\documentclass[%
reprint,
superscriptaddress,
nofootinbib,
nobibnotes,
amsmath,amssymb,
aps, prd
]{revtex4-2}

\usepackage{subcaption}
\usepackage{graphicx}
\usepackage{dcolumn}
\usepackage{bm}
\usepackage{hyperref}
\usepackage{amsmath} 
\usepackage{physics}
\usepackage{amssymb}
\usepackage{soul}
\usepackage{aas_macros}

\newcommand{\GA}{\alpha}
\newcommand{\GB}{\beta}
\newcommand{\GG}{\gamma}

\newcommand{\GC}{\psi}

\newcommand{\GP}{\phi}

\newcommand{\TUU}[3]{\tilde{#1}^{#2 #3}} 
\newcommand{\TDD}[3]{\tilde{#1}_{#2 #3}}

\newcommand{\fDu}[1]{\stackon[-0.3ex]{$D^{#1}$}{\kern-1.0ex\scalebox{0.7}{$\circ$}}}

\newcommand{\zD}{{\raise1.0ex\hbox{${}^{\ \circ}$}}\!\!\!\!\!D}

\usepackage{orcidlink}

\graphicspath{{./}{figures/}{figures/res_1e6_pngs/}{figures/res_1e4_pngs/}{figures/res_1e3_pngs/}{figures/res_1e2_pngs/}}

\begin{document}

\preprint{APS/123-QED}

\title{General-relativistic resistive-magnetohydrodynamics simulations of self-consistent magnetized rotating neutron stars}

\author{Patrick Chi-Kit \surname{Cheong}~\orcidlink{0000-0003-1449-3363}}
\email{patrick.cheong@berkeley.edu}
\affiliation{Department of Physics \& Astronomy, University of New Hampshire, 9 Library Way, Durham NH 03824, USA}
\affiliation{Center for Nonlinear Studies, Los Alamos National Laboratory, Los Alamos, NM 87545, USA}
\affiliation{Department of Physics, University of California, Berkeley, Berkeley, CA 94720, USA}

\author{Antonios \surname{Tsokaros}~\orcidlink{0000-0003-2242-8924}} 
\affiliation{Department of Physics, University of Illinois Urbana-Champaign, Urbana, IL 61801, USA}
\affiliation{National Center for Supercomputing Applications, University of Illinois Urbana-Champaign, Urbana, IL 61801, USA}
\affiliation{Research Center for Astronomy and Applied Mathematics, Academy of Athens, Athens 11527, Greece}

\author{Milton \surname{Ruiz}~\orcidlink{0000-0002-7532-4144}}
\affiliation{Departamento de Astronom\'{\i}a y Astrof\'{\i}sica, Universitat de Val\'encia, Dr. Moliner 50, 46100, Burjassot (Val\`encia), Spain}

\author{Fabrizio \surname{Venturi}~\orcidlink{0009-0003-5034-1073}}
\affiliation{Departamento de Astronom\'{\i}a y Astrof\'{\i}sica, Universitat de Val\'encia, Dr. Moliner 50, 46100, Burjassot (Val\`encia), Spain}

\author{Juno Chun Lung \surname{Chan}~\orcidlink{0000-0002-3377-4737}}
\affiliation{Niels Bohr International Academy, Niels Bohr Institute, Blegdamsvej 17, 2100 Copenhagen, Denmark}

\author{Anson Ka Long \surname{Yip}~\orcidlink{0009-0008-8501-3535}}
\affiliation{Department of Physics, The Chinese University of Hong Kong, Shatin, N.T., Hong Kong}

\author{K\=oji \surname{Ury\=u}~\orcidlink{0000-0002-7473-3587}}
\affiliation{Department of Physics, University of the Ryukyus, Senbaru, Nishihara, Okinawa 903-0213, Japan}

\date{\today}

\begin{abstract}
We present the first general-relativistic \emph{resistive} magnetohydrodynamics simulations of self-consistent, rotating neutron stars with mixed poloidal and toroidal magnetic fields. 
Specifically, we investigate the role of resistivity in the dynamical evolution of neutron stars over a period of up to 100 ms and its effects on their quasi-equilibrium configurations. 
Our results demonstrate that resistivity can significantly influence the development of magnetohydrodynamic instabilities, resulting in markedly different magnetic field geometries. 
Additionally, resistivity suppresses the growth of these instabilities, leading to a reduction in the amplitude of emitted gravitational waves. 
Despite the variations in magnetic field geometries, the ratio of poloidal to toroidal field energies remains consistently 9:1 throughout the simulations, for the models we investigated.
\end{abstract}

\maketitle


\section{\label{sec:intro}Introduction}
Neutron stars are among the most fascinating astrophysical objects in the universe. 
These stars embody some of the most extreme physical conditions, all contained within just a few kilometers in diameter. 
Along with their immense compactness and intense gravity, neutron stars also possess exceptionally strong, long-lived, large-scale magnetic fields.
The magnetic field strength of typical neutron stars, also known as pulsars~(see~e.g.~\cite{Manchester:2004bp}), could reach at least $\sim 10^{8-13}$ Gauss (G), orders of magnitude higher than in any other known object.
The strength of the neutron star magnetic field could be even more extreme and exotic, with values that reach $10^{15}$ G and beyond, up to 1000 times stronger than typical neutron stars. 
These ultra-strongly magnetized neutron stars, known as magnetars, are believed to be the sources of soft gamma-ray repeaters (SGRs) and anomalous X-ray pulsars~\citep{Mereghetti:2008je,2006csxs.book..547W}. 
Furthermore, the merger of two neutron stars can generate magnetic fields on the order of~$\sim 10^{16}-10^{17}\,\rm G$, which is even more extreme than the magnetic fields found in magnetars~\citep{Kiuchi2015a,Aguilera-Miret:2020dhz}.

The magnetic field structure of a neutron star is not yet fully understood, and gaining this understanding is crucial for interpreting expected multimessenger observations~\cite{LIGOScientific:2017ync}.
The most intuitive model of a magnetized neutron star which has been widely adopted assumes the star to be a huge two-pole magnet, meaning that they carry a large-scale dipolar magnetic field.
However, recent observational and theoretical evidence show that, the geometry of the neutron star magnetic field at the surface could be far from the conventional dipolar geometry, favoring instead a much more complicated configurations: multipolar magnetic field or an offset-dipole \citep{2019ApJ...887L..24M, 2019ApJ...887L..21R, 2019ApJ...887L..23B, 2020ApJ...889..165D}.
There is still no coherent theory of pulsar magnetosphere that would explain all the observational data from first-principles \citep{2018PhyU...61..353B,Contopoulos1999,Spitkovsky2006}.

The structure of the magnetosphere is influenced by the external magnetic field, which in turn is related to the internal field configuration of the neutron star.
Nevertheless, our current understanding of the internal magnetic field structure is limited due to the lack of direct observational evidence inside neutron stars.
Several efforts have been made by groups worldwide on understanding the magnetic field structure by means of quasiequilibria calculations
\citep{1995A&A...301..757B,
 2008PhRvD..78d4045K,
 2009ApJ...698..541K,
 2001ApJ...554..322C,
 2010MNRAS.401.2101Y,
 2012MNRAS.427.3406F,
 2015MNRAS.447.3785C,
 2016MNRAS.456.2937F,
 2014MNRAS.439.3541P,
 2015MNRAS.447.3278B,
 2017MNRAS.470.2469P}.
These studies are limited either in considering simple magnetic field configurations (typically pure poloidal or pure toroidal configurations) or by solving a simplified set of the Einstein-Maxwell system. 
In addition, a quasiequilibrium configuration does not imply stability. 
Therefore, a step forward involves:
i) computing quasiequilibria with more ``realistic'' magnetic field topologies, and 
ii) performing stability analysis in order to understand which of them can be physically permitted to represent a magnetized neutron star.
To address point (i), the numerical studies in~\cite{2014PhRvD..90j1501U, 2019PhRvD.100l3019U, Uryu:2023lgp} used the Compact Object CALculator (\texttt{COCAL}) code to compute fully general relativistic solutions for strongly magnetized, rapidly rotating compact stars. 
They achieved this by solving the full set of Einstein’s equations, coupled with Maxwell's and magnetohydrodynamic equations, under the assumptions of perfect conductivity, stationarity, and axisymmetry.
Strongly magnetized solutions associated with mixed poloidal and toroidal components of magnetic fields were successfully obtained in generic spacetimes. 

Magnetohydrodynamic (MHD) simulations are crucial for studying the stability of magnetic fields in magnetized neutron stars. 
Significant progress has been made by various groups in investigating the stability of these stars through general-relativistic dynamical simulations~\citep{2008PhRvD..78b4029K, 2011ApJ...736L...6C, 2011ApJ...735L..20L, 2012ApJ...760....1C, 2012arXiv1203.3590L, 2022MNRAS.511.3983S}.
In \cite{2008PhRvD..78b4029K} an initial toroidal magnetic field \cite{2008PhRvD..78d4045K} is evolved in full general relativity but with axisymmetry, while in \cite{2011ApJ...736L...6C, 2011ApJ...735L..20L, 2012ApJ...760....1C} purely poloidal magnetic fields \cite{1995A&A...301..757B}, have been evolved under the Cowling approximation, i.e., the Einstein field equations were not evolved but only the MHD equations on a fixed initial background. 
In \cite{2022PhRvL.128f1101T} full general-relativistic MHD simulations of self-consistent rotating neutron stars with ultra-strong mixed poloidal and toroidal magnetic fields~\citep{2019PhRvD.100l3019U} have been carried out for the first time.
Despite the fact that these models have been evolved for $\sim 10-20$ Alfv\'en timescales (as estimated from the initial conditions), these simulations were not long enough to further investigate the quasiequilibrium state of the magnetic fields. 

Although the ideal MHD approximation has been extensively used to model neutron stars, going beyond this approach is required for more accurate and realistic modeling of the plasma.
Important physical processes, such as dissipation and magnetic reconnection \cite{Pontin2022}, cannot be captured if one assumes vanishing electrical resistivity.
Magnetic reconnection can change the magnetic field's topology, and convert magnetic energy into other forms of energy such as heat and kinetic energy.
These processes, although usually occurring at very small length-scales, could significantly affect the large scale dynamics of the plasma, and hence the dynamical evolution of magnetized neutron stars. 
Although in practice, ideal MHD simulations exhibit non-zero numerically induced resistivity~\citep{Rembiasz_2017,komissarov2024splittingmethodnumericalrelativistic}, this effective resistivity is not well-controlled and cannot be adjusted as a physical parameter. To properly investigate the effects of resistivity, resistive MHD codes are required.
Several research groups have developed numerical codes to investigate the effects of resistive magnetohydrodynamics in a range of scenarios~\cite{Komissarov2007,Dumbser2009,Zenitani2010,2009MNRAS.394.1727P,Takamoto2011b,Bucciantini2012,PhysRevLett.111.061105,2013PhRvD..88d4020D,Palenzuela2013b,2013MNRAS.431.1853P,Ponce2014,2015PhRvD..92h4064D}.

In this work, we investigate the role of resistivity in the dynamical evolution of quasi-equilibrium configurations of neutron stars. 
To achieve this, we perform general-relativistic resistive MHD simulations of self-consistent, magnetized neutron star equilibria with ultra-strong, mixed magnetic fields. 
The simulations are run for up to $100\,\rm ms$ (approximately 100-200 initial Alfv\'en timescales), which is about 20 times longer than those conducted in~\cite{2022PhRvL.128f1101T}, allowing us to explore the potential final quasiequilibrium state of the magnetized neutron star.

The paper is organised as follows.
In Section~\ref{sec:methods} we outline the numerical methods we used in this work, along with the 
initial data and a suite of diagnostics used to verify the reliability of our numerical calculations.
We present our results in Section~\ref{sec:results}, and summarise our findings and 
conclusions in Section~\ref{sec:conclusions}.
Unless explicitly stated, we work in geometrized {Heaviside}-Lorentz units, for which the speed of light $c$, gravitational constant $G$, solar mass $M_{\odot}$, and the vacuum permittivity $\epsilon_0$ are all equal to one ($c = G = M_{\odot} = \epsilon_0 = 1$). 
Greek indices, running from 0 to 3, are used for spacetime objects, while Roman indices, running from 1 to 3, are used for spatial ones.

\section{\label{sec:methods}Methods}

\subsection{\label{sec:id}Initial profiles}
As initial data we use model ``A2'' from \citep{2022PhRvL.128f1101T}, which proved to be the most stable (in ideal GRMHD) investigated in that work. 
Table~\ref{tab:table_ID} provides a summary of its key physical properties.
This stationary and axisymmetric magnetar, with a polytropic $\Gamma=2$ equation of state, was constructed using the magnetized rotating neutron star libraries of the \texttt{COCAL} code \citep{2014PhRvD..90j1501U, 2019PhRvD.100l3019U} as a solution of the system of Einstein's field equations, Maxwell's equations, and the ideal MHD equations.

\begin{table*}
\caption{Key physical properties of the evolved model. 
We list the central rest-mass density, the gravitational mass, the period, the polar-to-equatorial-radio, the kinetic-to-gravitational-energy, the magnetic-to-gravitational-energy ratios, 
(including the toroidal and poloidal magnetic energies), the dynamical timescale ($1/\sqrt{\rho}$), and the Alfv\'en timescale.} 
\label{tab:table_ID}
\begin{tabular}{cccccccccc}
\hline\hline
$\rho_{c}$ & $M$  & $P/M$  & $R_p/R_e$ & $\mathcal{T}/|\mathcal{W}|$ & $\mathcal{M}/|\mathcal{W}|$  &  
$\mathcal{M}_{\rm tor}/|\mathcal{W}|$ & $\mathcal{M}_{\rm pol}/|\mathcal{W}|$ & $t_d/M$ & $t_A/M$   \\  
$10^{15.03}$ & $1.37$ & $169.3$ & $0.7$   & $10^{-1.31}$ & $10^{-1.79}$  & $10^{-3.10}$  & $10^{-1.82}$   & $18$ & $70$  \\
\hline\hline
\end{tabular}
\end{table*}

In summary, the Einstein equations under certain assumptions, can be written as a set of elliptic (Poisson-type) equations for $\{\GA,\GB^i,\GC,h_{ij}\}$, where $\GC$, $\GA$, $\GB^i$, and $h_{ij}$ are the conformal factor, lapse, shift vector, and the non-flat part of the conformal geometry ($\TDD{\GG}{i}{j}=f_{ij}+h_{ij}$). 
For the gauge conditions we use maximal slicing $K=0$, and the Dirac gauge $\zD_b\TUU{\GG}{a}{b}=0$, where $\zD_a$ is the covariant spatial derivative with respect to the flat metric $f_{ij}$. 
The 3+1 decomposition of Maxwell’s equations leads to four elliptic equations for the spatial components of the electromagnetic 1-form $A_\GA$, subject to the Coulomb gauge $\zD^i A_i=0$. 
The ideal MHD condition implies that surfaces of constant $A_t$ and $A_\GP$ coincide, and therefore these variables are functions of a single master potential, which is taken to be $A_\GP$ itself.
The first integrals of the MHD-Euler equations and the ideal MHD conditions become relations to determine the specific enthalpy $h$, the components of 4-velocity $u^t$ and $u^\phi$, and the components of the current $j^\alpha$. 
The latter involves the following arbitrary functions of the potential $A_\phi$, which are chosen to be
\begin{eqnarray}
\Lambda(A_\GP) & = & -\Lambda_0 \Xi(A_\GP) -\Lambda_1 A_\GP - \mathcal{E},  \label{eq:int1} \\
A_t(A_\GP)  & = & -\Omega_c A_\GP + C_e,  \label{eq:int2} \\
{[\sqrt{-g} \Lambda_\GP]} (A_\GP) & = & \Lambda_{\GP 0} \Xi(A_\GP). \label{eq:int3} 
\end{eqnarray}
In equations~\eqref{eq:int1}-\eqref{eq:int3}, $\Lambda_0, \Lambda_1$, and $\Lambda_{\GP 0}$ are input parameters that control the poloidal and toroidal magnetic field strength, while constants $\mathcal{E}$ and $\Omega_c$ are determined during the iteration procedure. 
The former represents the injection energy \cite{Friedman2012}, while the latter is the constant angular velocity of the magnetar. 
Constant $C_e$ controls the net charge of the star, which in our case is zero. 
$\Xi(A_\GP)$ is an integral of the ``sigmoid'' function \cite{2019PhRvD.100l3019U} which is used where $A_\GP$ varies on the fluid support, and its derivative is written
\begin{equation}
\Xi'(A_\phi)
= \frac12\left[\,\tanh\left(\frac1{b}
\frac{A_\phi-A_{\phi, \rm S}^{\rm max}}{A_\phi^{\rm max} - A_{\phi, \rm S}^{\rm max}}
-c\right)+1\,\right], 
\label{eq:xi}    
\end{equation}
where $A_\phi^{\rm max}$ is the maximum value of $A_\phi$, and $A_{\phi, \rm S}^{\rm max}$ is the maximum value of $A_\phi$ at the stellar surface.
Parameters $b,c \in [0,1]$ control the width and the position of the sigmoid.
Therefore, $\Xi'(A_\phi)$ vary from zero to one in the interval $A_\phi \in [A_{\phi, \rm S}^{\rm max}, A_\phi^{\rm max}]$. 
This guarantees that the current and the toroidal magnetic field are confined in the star, and the components of electromagnetic fields extend continuously into the exterior vacuum region.
Together with $\Lambda_0, \Lambda_1$, and $\Lambda_{\GP 0}$ the parameters $b$ and $c$ for the evolved model, are reported in Table~\ref{tab:param}. 

\begin{figure*}
	\centering
	\includegraphics[width=\textwidth, angle=0]{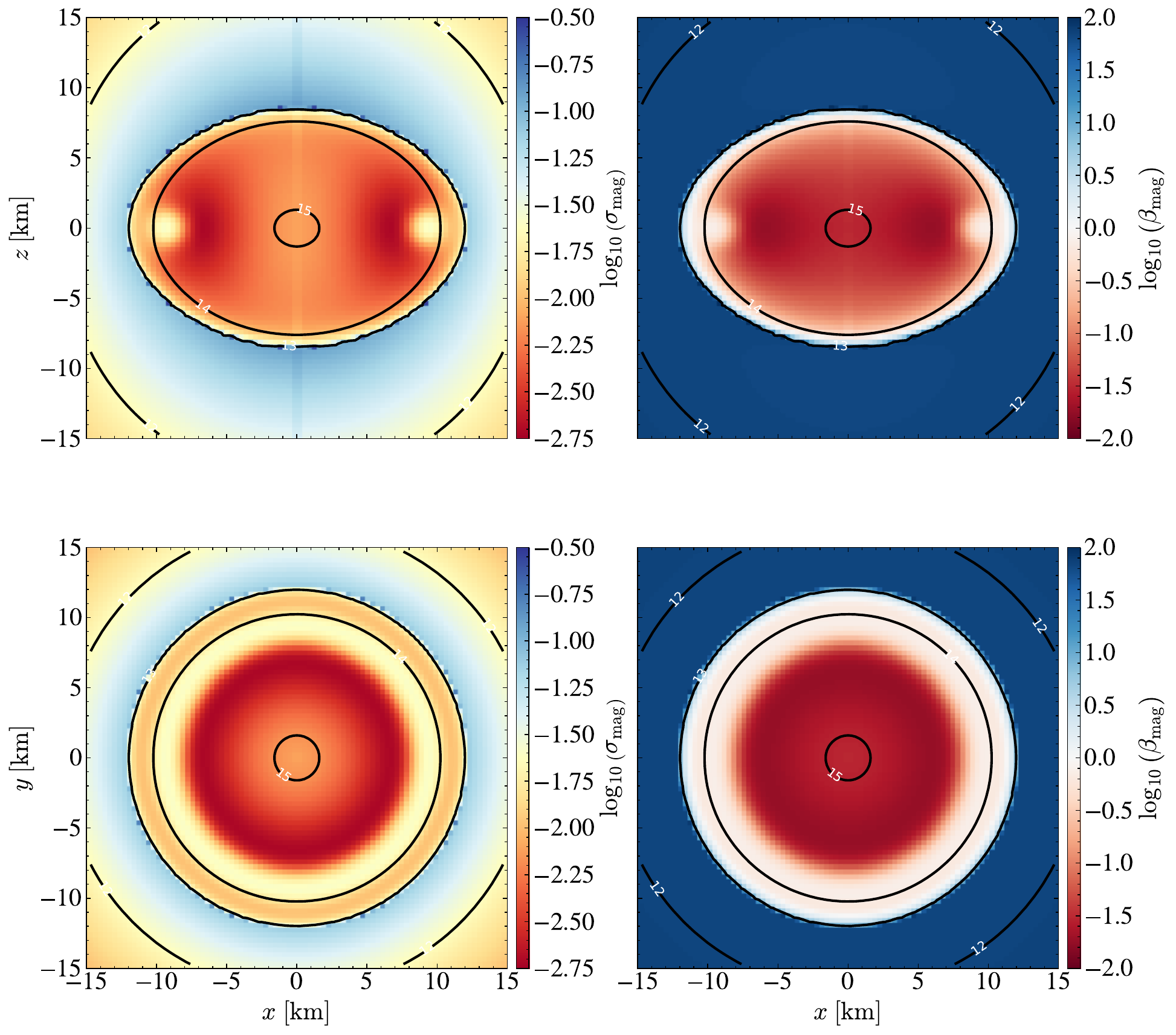}
	\caption{
		The logarithmic plasma sigma $\sigma_{\rm mag}$ (\emph{left column}) and the magnetic-to-gas pressure ratio $\beta_{\rm mag}$ (\emph{right column}) profiles on the $xz$-plane (\emph{top panels}) and the $xy$-plane (\emph{bottom panels}) at the beginning of the simulations.
		Black contours correspond to the rest mass densities of $10^{15, 14, 13, 12}~{\rm g \cdot cm^{-3}}$.
  		\label{fig:sigma_beta}
		}
\end{figure*}

\subsection{\label{sec:sim}Evolutions}

General-relativistic \emph{resistive} MHD equations are solved with the \texttt{Gmunu} code~\citep{2020CQGra..37n5015C, 2021MNRAS.508.2279C, 2022ApJS..261...22C}.
In particular, the full Maxwell and hydrodynamic equations are solved, where the coupling between electromagnetic fields and fluid are determined by the electric current.
In this work, we choose Ohm's law by following~\cite{2013MNRAS.428...71B}
\begin{equation}
	J^i = \rho_e v^i + \frac{W}{\eta} \left[ E^i + \epsilon^{ijk}v_jB_k - \left( E^j v_j \right) v^i\right],
\end{equation}
where $J^i$ is the 3-current, $\rho_e$ is the charge density, $v^i$ is the velocity, $\eta$ is the resistivity, $W$ is the Lorentz factor, $\epsilon^{ijk}$ is the spatial Levi-Civita pseudo-tensor, $E^i$ and $B^i$ are the electric and magnetic fields, respectively.
Divergence-free condition is preserved by using staggered-meshed constrained transport \citep{1988ApJ...332..659E}.
For the details of the implementations of the resistive MHD in \texttt{Gmunu}, we refer readers to \citep{2022ApJS..261...22C}.

We explore four different resistivities, namely, $\eta \in \{10^{-6}, 10^{-4}, 10^{-3}, 10^{-2}\}$.
The Ohmic diffusion timescale can be estimated as (e.g. \cite{2015PhRvD..92h4064D})
 \begin{equation} \label{eq:tdis}
     t_{\rm dis} \approx \frac{L^2}{\eta}  \, ,
 \end{equation}
where $L$ is the typical length scale of the system, which here we take as 10 km.
For the chosen values of resistivity, the orders of magnitude of the corresponding Ohmic diffusion timescales, $t_{\rm dis}$, are $\{10^{5}, 10^{3}, 10^{2}, 10^{1}\} ~\rm{ms}$.
Indeed, the typical value of resistivity is unknown (see e.g. \citep{2021PhRvD.103d3022S}).
We choose the values that cover both dynamical and secular timescales.
The resistivity is set to be uniform everywhere in the computational domain. 
In comparison, the initial Alfv\'en timescale\footnote{
A more accurate estimate of the Alfv\'en timescale during the initial stages of ideal MHD evolution can be found in the Supplement of~\cite{2022PhRvL.128f1101T}, and it roughly aligns  with this timescale.}
\begin{equation}
     t_{\rm A} \approx \frac{R_e \sqrt{\rho}}{B} = 0.5\ {\rm ms}\, ,
     \label{eq:tA}
 \end{equation}
where B is the value of the magnetic field at the neutron star centre, and $R_e$ the equatorial radius, is comparable to the dynamical timescale $t_d\approx 0.1$ ms. 

For the evolutions, we use a $\Gamma$-law equation of state $P_{\rm gas} = \left(\Gamma-1\right) \rho \epsilon$, where $\rho$ is the rest mass density and $\epsilon$ is the specific internal energy. 
We set $\Gamma=2$ to match the initial model.
\begin{figure}
	\centering
	\includegraphics[width=\columnwidth, angle=0]{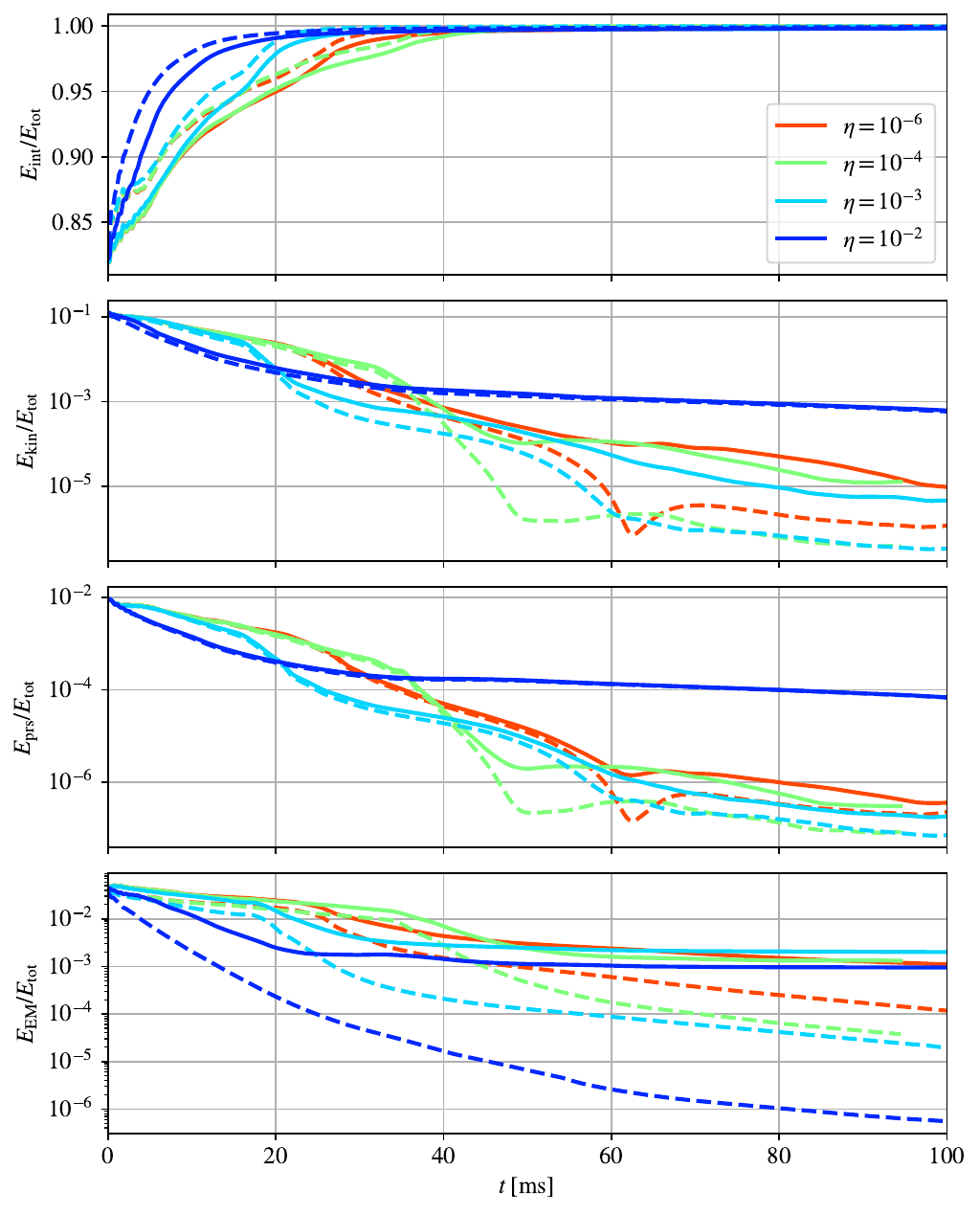}
	\caption{
			Time evolution of the internal energy $E_{\rm int}/E_{\rm tot}$ (top panel), kinetic energy $E_{\rm kin}/E_{\rm tot}$ (second from the top panel), pressure contribution $E_{\rm prs}/E_{\rm tot}$ (third from the top panel), and electromagnetic energy $E_{\rm EM}/E_{\rm tot}$ (bottom panel) for the strongly magnetized, rapidly rotating neutron star model in Table~\ref{tab:table_ID} with different resistivities $\eta$.
			$E_{\rm tot}$ is the total energy of the system. 
			Dashed lines are calculated considering only the interior of the star.
		\label{fig:all_energy}
		}
\end{figure}

\begin{figure}
	\centering
	\includegraphics[width=\columnwidth, angle=0]{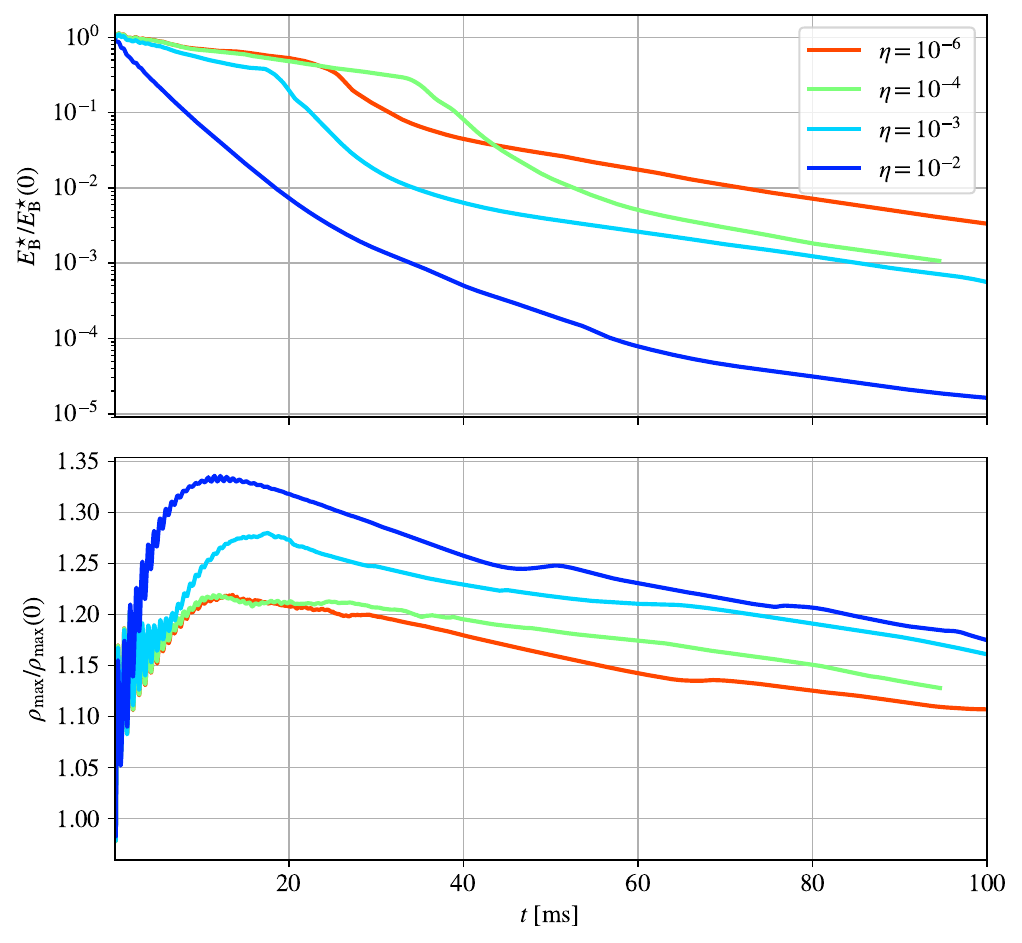}
	\caption{
		Time evolution of the internal magnetic energy $E^{\star}_B$ (upper panel) and the maximum rest-mass density $\rho_{\max}$ (lower panel) of the strongly magnetized, rapidly rotating neutron star model in Table~\ref{tab:table_ID} with different resistivity $\eta$.
		\label{fig:rhomax}
		}
\end{figure}
\begin{figure}
	\centering
	\includegraphics[width=\columnwidth, angle=0]{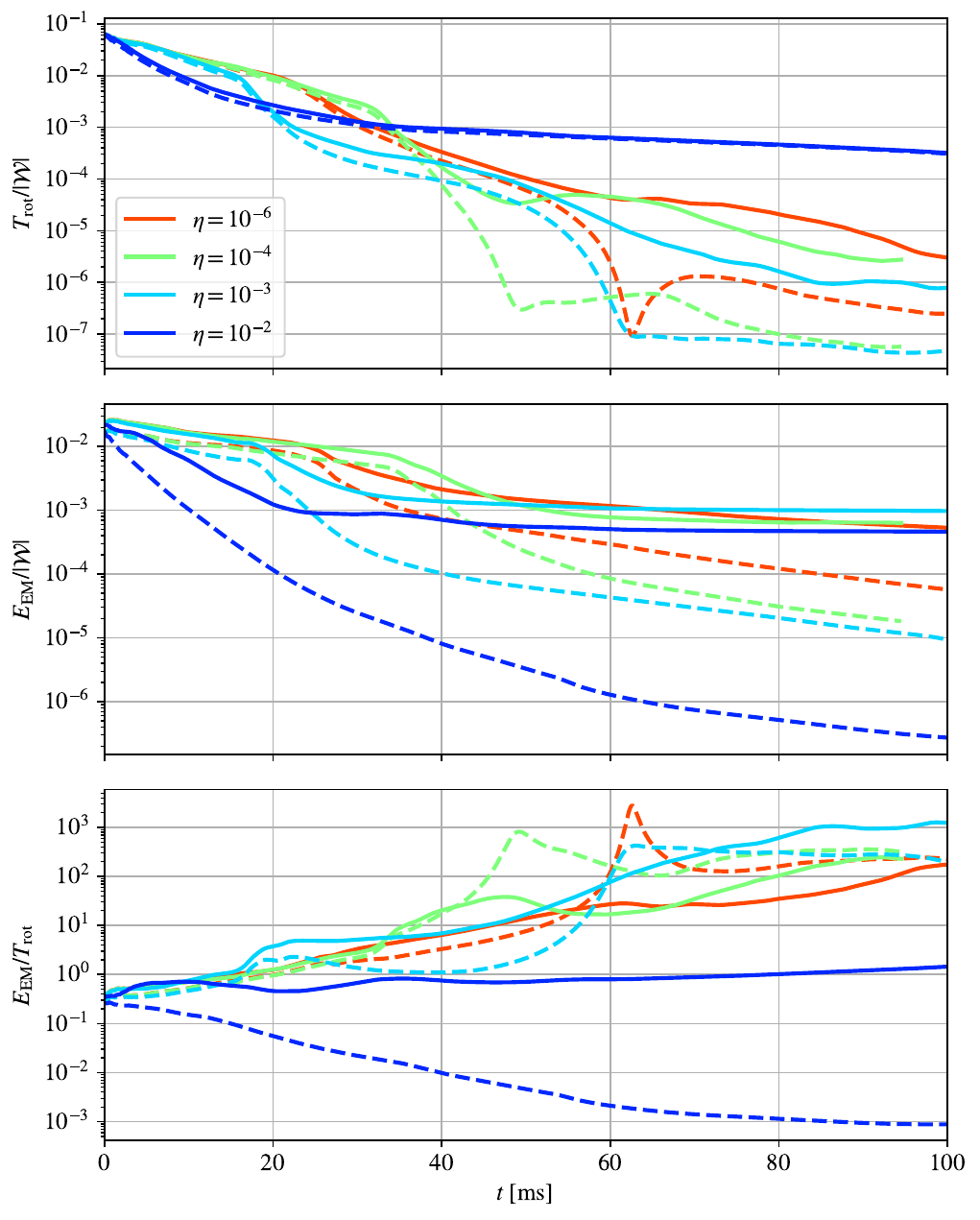}
	\caption{
		Time evolution of the rotational kinetic-to-gravitational-energy ratio $T_{\rm rot} / \lvert \mathcal{W} \rvert$ (top panel), electromagnetic-to-gravitational-energy ratio $E_{\rm EM} / \lvert \mathcal{W} \rvert$ (middle panel), and their ratio $E_{\rm{EM}}/T_{\rm{rot}}$ (bottom panel) for different resistivities $\eta$.
		Dashed lines are calculated considering only the interior of the star.
  		\label{fig:beta_and_Eem_over_T}
		}
\end{figure}

All the simulations here has been performed in Cartesian coordinates $(x, y, z)$ without imposing any symmetries.
The computational domain covers $[-120,120]$ along $x$, $y$ and $z$, with the resolution $N_x \times N_y \times N_z = 128 \times 128 \times 128$ and allowing 4 adaptive mesh refinement (AMR) level.
The finest grid size at the centre of the star is $\Delta x = \Delta y = \Delta z \approx 346 ~\rm{m}$.
The refinement is fixed after the initialization, since we do not expect the stars to expand significantly.
Our simulations adopt Harten, Lax and van Leer (HLL) approximated Riemann solver~\citep{harten1983upstream}, 3rd-order reconstruction piecewise parabolic method (PPM)~\citep{1984JCoPh..54..174C}. 
Implicit-explicit Runge-Kutta scheme IMEXCB3a~\citep{2015JCoPh.286..172C} is used to deal with the stiff terms in the evolution equations for small resistivity.

At the beginning of the simulations, we impose a low and variable density magnetosphere with the magnetic-to-gas pressure ratio $\beta_{\rm mag} \equiv P_{\rm mag} / P_{\rm gas} = 10^{2}$ everywhere outside the star by following \citep{2018PhRvD..98l3017R}.
This approach has been used in~\citep{2015ApJ...806L..14P, 2016ApJ...824L...6R, 2018PhRvD..98l3017R, 2020PhRvD.102j4022R} to reliably evolve the magnetic field in magnetic-pressure dominant environments. 
During the evolution, points will be treated as the ``atmosphere'' when their rest-mass density $\rho$ drops below the threshold value, $\rho_{\rm thr}$. 
In this case, we set the rest-mass density to be $0.99 \rho_{\rm thr}$ and the velocity to be zero (i.e. $v^i = 0$). 
The threshold value $\rho_{\rm thr}$ is set to be 10 orders of magnitude smaller than the initial maximum density, i.e. $\rho_{\rm thr} = 10^{-10} \times \rho_{\rm max}(t=0)$. 
The initial configuration of the plasma sigma $\sigma_{\rm mag} \equiv 2 P_{\rm mag} / \rho$ and the magnetic-to-gas pressure ratio $\beta_{\rm mag}$ at the beginning of the simulations are shown in Fig.~\ref{fig:sigma_beta}.

All models are evolved in a dynamical spacetime under the conformally-flat approximation. 
Although the initial data from \texttt{COCAL} is fully general relativistic, the conformally-flat approximation used in \texttt{Gmunu} is sufficient for the purpose of this study. 
For the magnetar model that we consider here, whose compactness is $\sim 0.14$, the off-diagonal components of the 3-metric $\gamma_{ij}$ are very small compared to the diagonal components, making the initial data nearly conformally flat. 
Studies have shown that there is, at most, a few percent difference between fully general relativistic and conformally flat rotating equilibrium models~\citep{1996PhRvD..53.5533C, 2014GReGr..46.1800I, 2021MNRAS.503..850I, 2022MNRAS.510.2948I}, and such quasispherical equilibria are very likely to remain stable even with high angular momentum~\citep{2024PhRvD.110d3015C}. 
Therefore, in this work, the off-diagonal components in the 3-metric $\gamma_{ij}$ in the initial data are simply ignored. 
To verify this approach, we perform a zero-resistivity simulation and compare it with the one reported in \citep{2022PhRvL.128f1101T} which employs the well-known fully general relativistic \texttt{IllinoisGRMHD}~(see e.g.~\citep{Ruiz:2021gsv}). 
This comparison is presented in Appendix~\ref{sec:code_tests}. 

\begin{table}
\caption{Parameters of functions in the integrability conditions equations~\eqref{eq:int1}-\eqref{eq:int3} and equation~\eqref{eq:xi} for the evolved magnetar.}
\label{tab:param}
\begin{tabular}{ccccccc}
\hline
$\phantom{-}\Lambda_0$ & $\Lambda_1$ & $\Lambda_{\phi0}$ & $b$ & $c$  \\
\hline
$-0.6$ & $0.3$ & $1.1$ & $0.2$ & $0.5$   \\
\hline
\end{tabular}
\end{table}
\subsection{\label{sec:diagnostics}Diagnostics}
The rest, proper, and gravitational (ADM) masses are computed as \cite{2008PhRvD..78b4029K}: 
\begin{eqnarray}
	M_{\rm rest} &=& \int \rho W \psi^6 \dd{x}^3, \label{eq:Mrest}\\
	M_{\rm proper} &=& \int \rho W \left( 1 + \varepsilon \right) \psi^6 \dd{x}^3, \label{eq:Mpr} \\
	M_{\rm ADM} &=& \int \left[ \rho_H + \frac{K^{ij}K_{ij}}{16 \pi}\right] \psi^5 \dd{x}^3, \label{eq:Madm}
\end{eqnarray}
where $\varepsilon$ is the fluid specific internal energy, $\psi$ the conformal factor, $K^{ij}$ is the extrinsic curvature, and
\begin{equation}
 \rho_H= T_{\alpha\beta}n^\alpha n^\beta =  \rho h W^2 - P + \frac{1}{2}(E^2+B^2) . 
\label{eq:rhoH}
\end{equation}
Here, the total stress-energy tensor $T_{\alpha\beta}$  is the sum of the stress-energy tensor for a perfect fluid and the stress-energy tensor for the electromagnetic field, $n^\alpha$ is the normal to the hypersurface, $W$ is the Lorentz factor, $P = P_{\rm gas}$ is the pressure of the fluid, $h=1+\varepsilon +P/\rho$ is the specific enthalpy, $E^2=E^iE_i$, $B^2=B^iB_i$, and $E^i$, $B^i$ are the purely spatial electric, magnetic fields with respect to the normal observer. 
Therefore, the support of the volume integral \eqref{eq:Madm} is non-compact.

The conserved energy density $\tau$, which is the ADM energy (equation~\eqref{eq:rhoH}) but without the rest mass contribution (i.e. $\tau = \rho_H - \rho W$), is being evolved in \texttt{Gmunu} \cite{2022ApJS..261...22C}.
This conserved energy density can be expanded as
\begin{equation}\label{eq:tau}
\begin{aligned}
	{\tau} = & \epsilon_{\rm int} + \epsilon_{\rm kin} + \epsilon_{\rm prs} + \epsilon_{\rm EM}
\end{aligned},
\end{equation}
where $\epsilon_{\rm int}$, $\epsilon_{\rm kin}$, $\epsilon_{\rm prs}$, and $\epsilon_{\rm EM}$ are the internal, kinetic, pressure contribution, and electromagnetic energy densities.
They can be obtained by
\begin{eqnarray}
	\epsilon_{\rm int} &=& \rho W^2 \varepsilon, \\
	\epsilon_{\rm kin} &=& \rho W \left( W - 1 \right), \\
	\epsilon_{\rm prs} &=& P \left( W^2 - 1 \right), \\
	\epsilon_{\rm EM}  &=& \frac{1}{2}\left(B^2 + E^2\right).
\end{eqnarray}
We can then obtain different energies by
\begin{eqnarray}
	E_{\rm int} &=& \int \epsilon_{\rm int} \sqrt{\gamma} \dd{x}^3, \\
	E_{\rm kin} &=& \int \epsilon_{\rm kin} \sqrt{\gamma} \dd{x}^3, \\
	E_{\rm prs} &=& \int \epsilon_{\rm prs} \sqrt{\gamma} \dd{x}^3, \\
	E_{\rm EM}  &=& \int \epsilon_{\rm EM} \sqrt{\gamma} \dd{x}^3,
\end{eqnarray}
where $\gamma$ is the determinant of the 3-metric $\gamma_{ij}$. 
The total energy can be obtained either by summing up all the energies (i.e. $E_{\rm tot} =E_{\rm int} +E_{\rm kin} +E_{\rm prs} +E_{\rm EM}$), or by integrating all the total conserved energy density 
\begin{equation}
	E_{\rm tot} = \int \tau \sqrt{\gamma} \dd{x}^3.
\end{equation}

The rotational kinetic energy is given by
\begin{align}
	T_{\rm rot} &= \frac{1}{2} \int \Omega \left( x S_y - y S_x \right) \sqrt{\gamma} \dd{x}^3,
\end{align}
where $\Omega$ is angular velocity, while $S_i = \rho W^2 h v_i$ are the conserved matter momenta.
The angular velocity $\Omega$ is given by
\begin{align}
	\Omega &= \frac{x \hat{v}^{y} - y\hat{v}^{x} }{x^2+y^2},
\end{align}
where $\hat{v}^i = \alpha v^i - \beta^i$.
The gravitational binding energy $\mathcal{W}$ is defined as
\begin{align}
	\mathcal{W} &= M_{\rm ADM} - M_{\rm proper} - T_{\rm rot}.
\end{align}

Gravitational waves are extracted via the quadrupole formula~\cite{2001ApJ...548..919S}:
\begin{align}
	h^{\rm p}_{+} =& \frac{1}{d_{\rm obs}} \left( \ddot{I}_{xx} - \ddot{I}_{yy} \right), \\
	h^{\rm p}_{\times} =& \frac{1}{d_{\rm obs}} \left( 2\ddot{I}_{xy} \right), \\
	h^{\rm e}_{+} =& \frac{1}{d_{\rm obs}} \left( \ddot{I}_{zz} - \ddot{I}_{yy} \right), \\
	h^{\rm e}_{\times} =& \frac{1}{d_{\rm obs}} \left( - 2\ddot{I}_{yz} \right), 
\end{align}
where $h^{\rm p}$ and $h^{\rm e}$ are the gravitational waves strains observed on the polar axis and the equatorial plane, respectively.
Here~$d_{\rm obs}$ is the distance from the source, which is assumed to be 10~Mpc, while
$\ddot{I}_{ij}$ is the second time derivative of quadrupole moment $I_{ij}$.
In practice, we obtain $\ddot{I}_{ij}$ by taking time derivatives of the first time derivative of quadrupole moment $\dot{I}_{ij}$ via post-processing~\cite{2001ApJ...548..919S}.
The first time derivative of quadrupole moment is calculated directly from the simulation by
\begin{equation}
	\dot{I}_{ij} = \int \rho W \left( \hat{v}^i x^j + \hat{v}^j x^i \right)\sqrt{\gamma} \dd{x}^3.
\end{equation}
Note that, in the axisymmetric cases, only $h^{\rm e}_+$ is non-zero, while $h^{\rm p}_{+}$, $h^{\rm p}_{\times}$, $h^{\rm e}_{\times}$ all vanish.
Since the magnetar is axisymmetric, the gravitational wave strain $h^{\rm e}_+$ is expected to be much larger than others. 

To better focus on the interior of the neutron star, some diagnostics are computed only within the bulk of the star. 
In particular, for the region where the rest-mass density is higher than the neutron star surface density is regared as the bulk of the star.  
Here, the neutron star surface density is defined as
\begin{equation}
	\rho_{\rm surf} := 3 \times 10^{-2} \rho_{\rm max} \left(t=0 \right).
 \label{eq:rhosurf}
\end{equation}
We empirically find that this value is low enough to capture the high density regions and to visualize reliably the surface of neutron star (see below).
Different choices of the star's surface definition do not affect the diagnostics substantially, thus we use this value consistently throughout the paper, unless otherwise specified.

We calculate the total, toroidal, and poloidal magnetic energies within the interior of the neutron star according to (see~e.g.~\cite{2014MNRAS.439.3541P})
\begin{align}
	E^{\star}_{B} &= \frac{1}{2} \int_{\rho \geq \rho_{\rm surf}} B^i B_i \sqrt{\gamma} \dd{x}^3, \\
	E^{\star}_{B_{\rm tor}} &= \frac{1}{2} \int_{\rho \geq \rho_{\rm surf}} B^\phi B_\phi \sqrt{\gamma} \dd{x}^3, \\
	E^{\star}_{B_{\rm pol}} &= E^{\star}_{B} - E^{\star}_{B_{\rm tor}}.
\end{align}

To better understand the development and saturation of the instability of the star, we compute the volume-integrated azimuthal mode decomposition of a quantity $f$, which can be either the conserved rest-mass density $D := W \rho$ or the toroidal magnetic field $B^{\phi}$.
The volume-integrated azimuthal mode decomposition is defined as~(see~e.g.~\cite{Zurek1986})
\begin{equation}
	C^{\star}_{m}\left( f \right) = \int_{\rho \geq \rho_{\rm surf}} f \; e^{i m \phi} \; \sqrt{\gamma}\dd{x}^3,
\end{equation}
where $\phi \equiv \tan^{-1}\left(y/x\right)$ is the azimuthal angle and $m$ is the azimuthal number.

\section{\label{sec:results}Results}
\subsection{\label{sec:sum_of_evo}Summary of the evolutions}
The Ohmic dissipations of magnetic fields is one of the major effects due to resistivity.
We first investigate the magnetic energy dissipations. 
In Fig.~\ref{fig:all_energy} we show the time evolution of the internal energy, kinetic energy, the energy from the pressure contribution, and electromagnetic energy for different valued of the resistivity~$\eta$. 
Solid lines show the corresponding quantities when the integration is performed in the whole computational domain, while dashed lines
when the integration is performed within the neutron star surface, as defined by equation~\eqref{eq:rhosurf}.
In all cases, almost all the energy is converted into internal energy by the end of the simulations. 
Apart from the evolution with the highest resistivity, we can identify three epochs that are characterized by different decay rates. 
In the first epoch, when $t\lesssim 25, 35, 18$~ms for $\eta=10^{-6}, 10^{-4}, 10^{-3}$, respectively, the kinetic, pressure, and electromagnetic energies decay mildly. 
In the second epoch, when $t\lesssim 60, 50, 30$~ms correspondingly, a rapid decay of these energies is observed, which is followed by the third epoch where the decay rate decreases again, all the way to the end of our simulations. 
On the other hand, the evolution with the highest resistivity $\eta=10^{-2}$ is dominated by an initial rapid decrease of the kinetic, pressure, and electromagnetic energies, followed (at $t\sim 20$ ms) by a milder decay. 
At the same time, the internal energy shows a corresponding increase that results into heating up the star.

Fig.~\ref{fig:rhomax} shows the time evolution of the interior magnetic energy $E^{\star}_{B}$ and the maximum rest mass density $\rho_{\max}$ of our strongly magnetized rapidly rotating neutron star with different resistivity $\eta$. 
The evolution of the interior magnetic energy follows the same qualitative behavior as the total magnetic energy $E_{\rm EM}$ in Fig.~\ref{fig:all_energy}, although in the case of the highest resistivity the decay rate here is more uniform.
As the star loses magnetic pressure, it begins to contract, leading to an increase in the maximum rest-mass density.
The maximum rest-mass density in all cases decreases after $\sim 20~\rm{ms}$.
The larger the resistivity, the faster the decay of the magnetic energy, and the faster the increase of the maximum rest mass density.
The decay rates of the magnetic energy and maximum rest-mass density are about the same at the late times (i.e. $t \gtrsim 80~\rm{ms}$) despite different resistivity.

Spin down behavior is observed in all cases.
In Fig.~\ref{fig:beta_and_Eem_over_T} (top panel) the rotational kinetic energy over the gravitational binding energy is plotted. 
For all cases but for the highest resistivity one, the loss of rotational energy is almost constant and independent of the resistivity. 
For the case with $\eta=10^{-2}$, initially the decay is faster but after a certain time ($\sim 30$ ms) the decay becomes smaller. 
At the end of our simulations ($\sim 100$ ms) the highest resistivity case retains significant more kinetic energy than the other three cases (almost two orders of magnitude). 
Given the fact that the decay of electromagnetic energy is similar in all models (second panel), the ratio $E_{\rm{EM}}/T_{\rm{rot}}$ (bottom panel) is significantly different for $\eta=10^{-2}$ and results in an equipartition at $t\sim 100$ ms. 
On the contrary, in the cases with smaller resistivity, this ratio reaches values beyond 100, for $t\gtrsim 60$ ms. 
Notice that most of the electromagnetic energy at the end of the simulations is outside the star.

\begin{figure}
	\centering
	\includegraphics[scale=0.38]{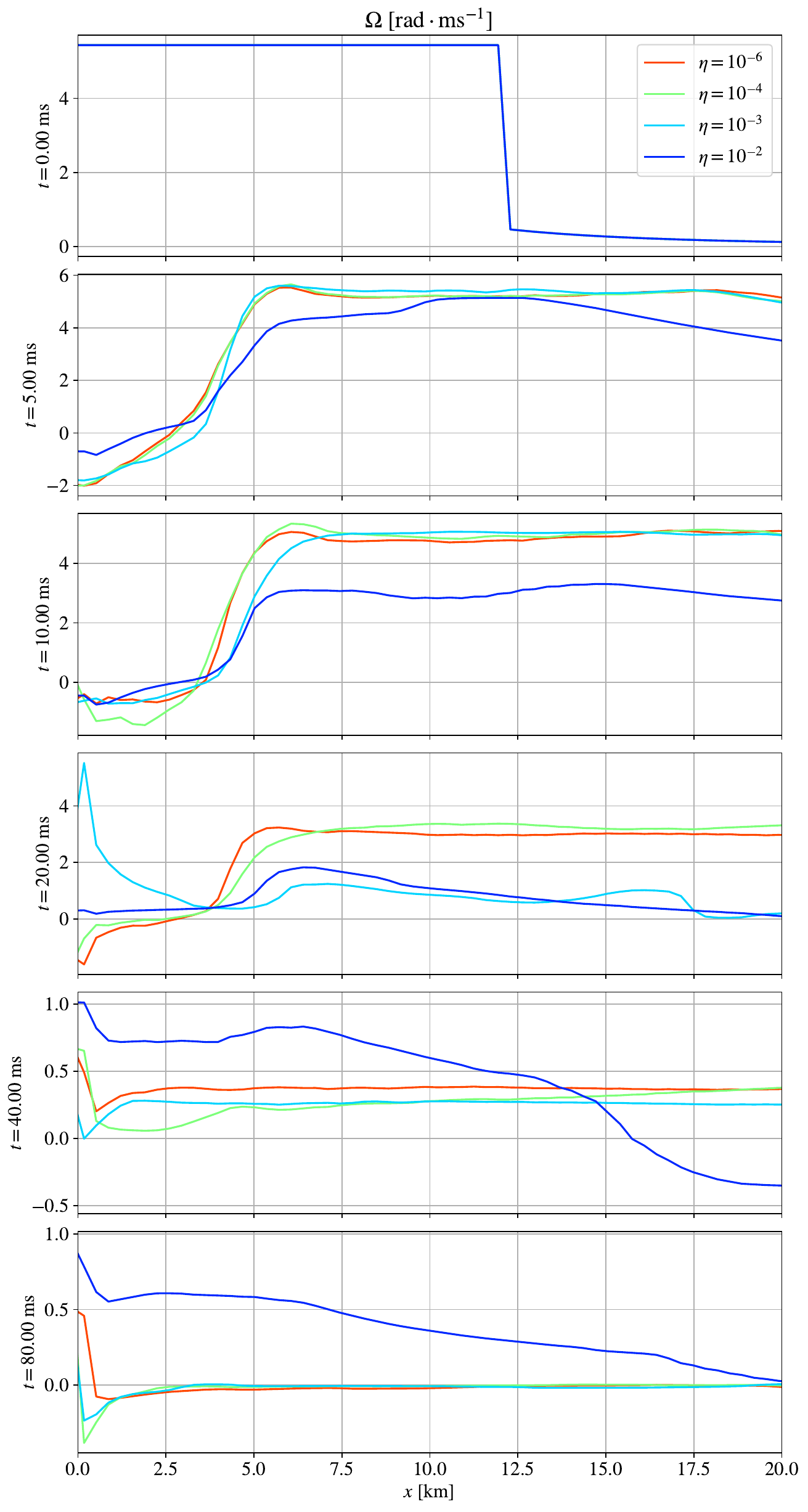} 
	\caption{
		Angular velocity profile $\Omega$ along the coordinate $x$-axis at various instances of the strongly magnetized and rapidly rotating neutron star model in Table~\ref{tab:table_ID} for different resistivities $\eta$.
		Different time instances are shown from the top to the bottom panels.	 
		\label{fig:combined_omega}
		}
\end{figure}

\begin{figure}
	\centering
	\includegraphics[scale=0.38]{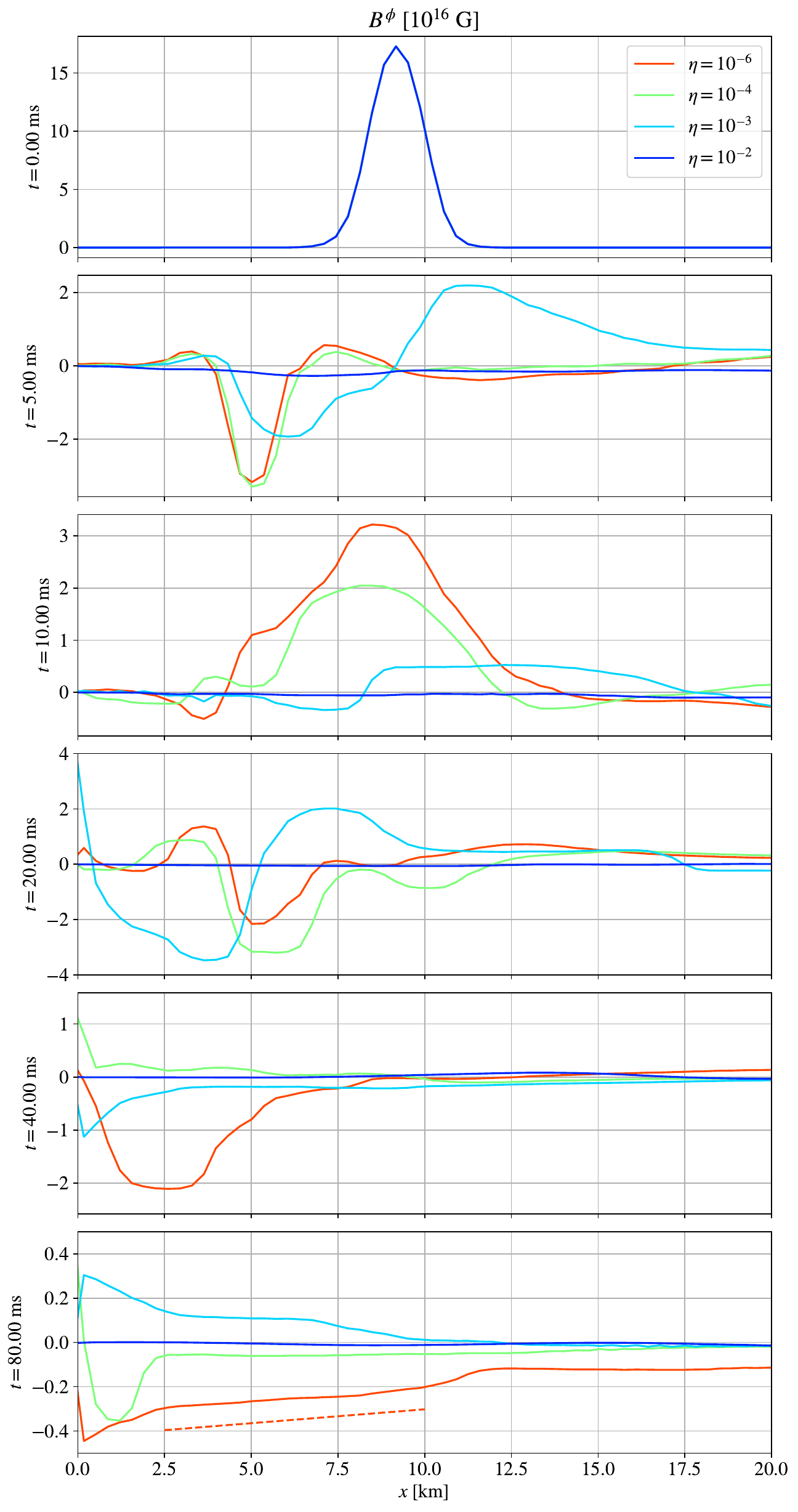}
 \caption{
		Toroidal magnetic field profile $B^{\phi}$ along the coordinate $x$-axis at various instances with different resistivities $\eta$. 	
		\label{fig:Bphi_profile}
		}
\end{figure}

The spin down behavior can also be seen in Fig.~\ref{fig:combined_omega} where the angular velocity profile of the neutron star along the coordinate $x$-axis is plotted at various instances~$\Omega\left(x,t\right)$ varying the resistivity~$\eta$.
The star is initially uniformly rotating with $\Omega \sim 5~{\rm rad \cdot ms^{-1}}$.
At the beginning, for $t \lesssim 5~\rm{ms}$, the angular velocity at the center of the star drops very quickly and fluctuates around zero (reminiscent of the behavior found in analytical models \cite{Shapiro00}), resulting in a differentially rotating neutron star.
This behavior agrees with the one reported in \cite{2022PhRvL.128f1101T} and is independent of resistivity.
Both in the highly resistive case with $\eta=10^{-2}$, as well as in the almost ideal MHD case with $\eta=10^{-6}$, the evolution of angular velocity profile is similar for $t\lesssim 10$ ms, while in later times the highly resistive case retains more of its rotational kinetic energy. 
It is probable, that further evolution will drain this kinetic energy even for the high resistivity case and a final profile similar to the other cases ($\eta=10^{-3}, 10^{-4}, 10^{-6}$) will be reached.
By the end of the simulations, all cases result in a slowly rotating star with a nearly spherical shape.

Fig.~\ref{fig:Bphi_profile} shows the toroidal magnetic field along the coordinate $x$-axis, at various time instances.
Initially, the toroidal magnetic field, with an order of magnitude $\sim 10^{17}~{\rm G}$, is concentrated just below the surface of the neutron star (top panel), but soon after the evolution starts, it becomes unstable and oscillating in direction irrespective of the resistivity. 
For the highest resistivity $\eta=10^{-2}$, the toroidal magnetic field disappears by the end of our simulations, while for the lowest resistivity (nearly ideal MHD) it reverses direction, becomes maximum near the centre and decays linearly (a red dashed linear function is shown in the bottom panel of Fig.~\ref{fig:Bphi_profile}).
It is important to note that Newtonian analysis of pure toroidal magnetic fields~\cite{Tayler1973} predicts that, for a star containing a toroidal field to remain stable against short timescale instabilities, the toroidal field must decay with cylindrical radius $\varpi=\sqrt{x^2 +y^2}$ at a rate at least as fast as $\varpi^2$. 
Rotation can be a stabilizing factor.

\begin{figure*}
	\centering
	\begin{subfigure}{0.32\textwidth}
		\includegraphics[width=\textwidth]{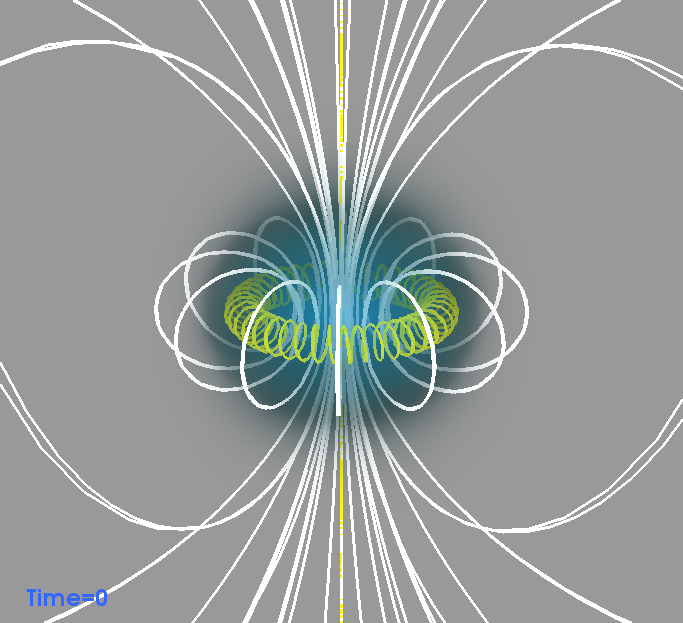}
		\caption{$t=0~{\rm ms}$}
	\end{subfigure}
	\hfill
	\begin{subfigure}{0.32\textwidth}
		\includegraphics[width=\textwidth]{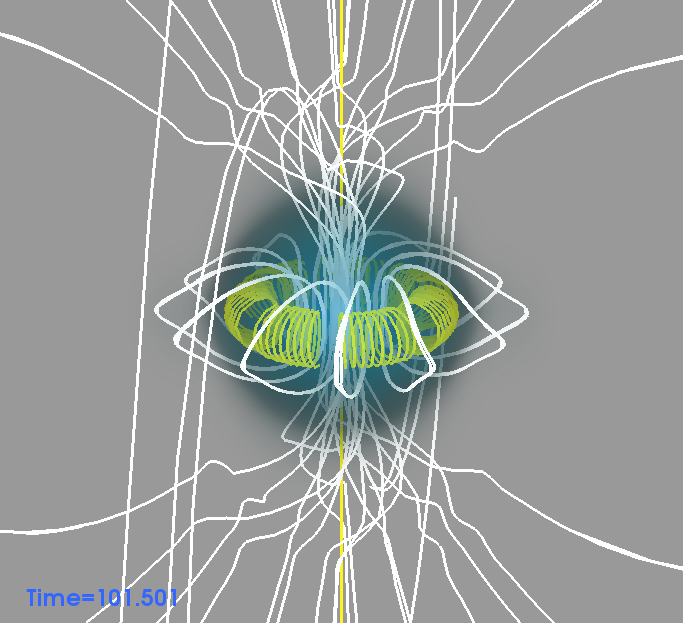}
		\caption{$t=0.5~{\rm ms}$}
	\end{subfigure}
	\hfill
	\begin{subfigure}{0.32\textwidth}
		\includegraphics[width=\textwidth]{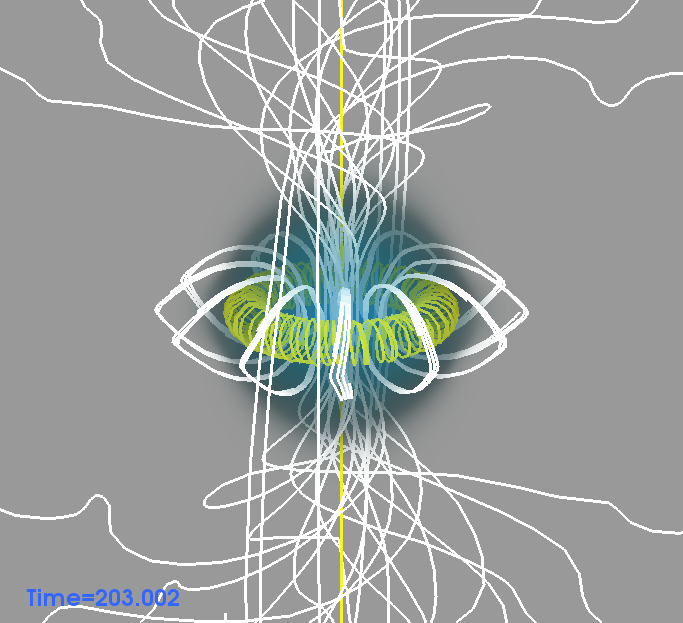}
		\caption{$t=1.0~{\rm ms}$}
	\end{subfigure}

	\begin{subfigure}{0.32\textwidth}
		\includegraphics[width=\textwidth]{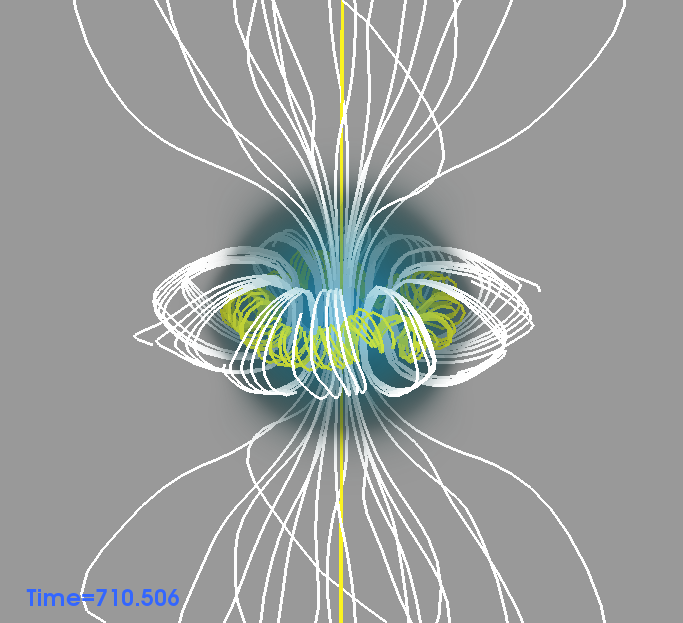}
		\caption{$t=3.5~{\rm ms}$}
	\end{subfigure}
	\hfill
	\begin{subfigure}{0.32\textwidth}
		\includegraphics[width=\textwidth]{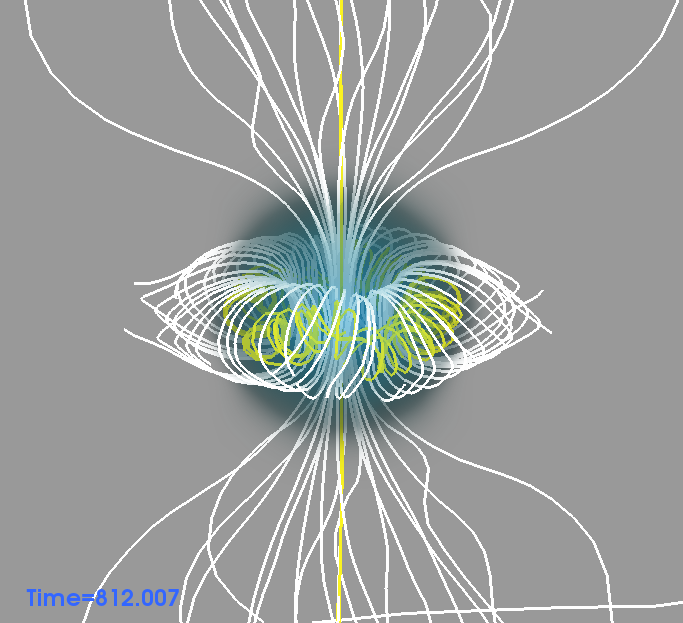}
		\caption{$t=4~{\rm ms}$}
	\end{subfigure}
	\hfill
	\begin{subfigure}{0.32\textwidth}
		\includegraphics[width=\textwidth]{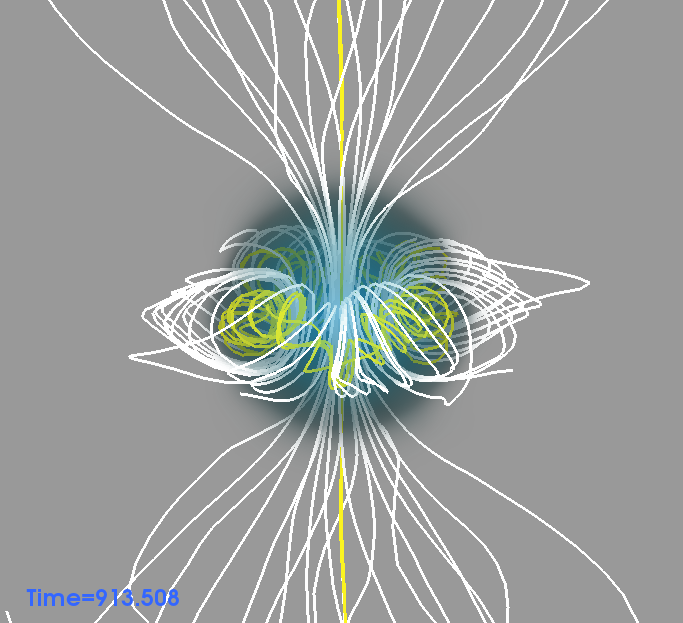}
		\caption{$t=4.5~{\rm ms}$}
	\end{subfigure}

	\begin{subfigure}{0.32\textwidth}
		\includegraphics[width=\textwidth]{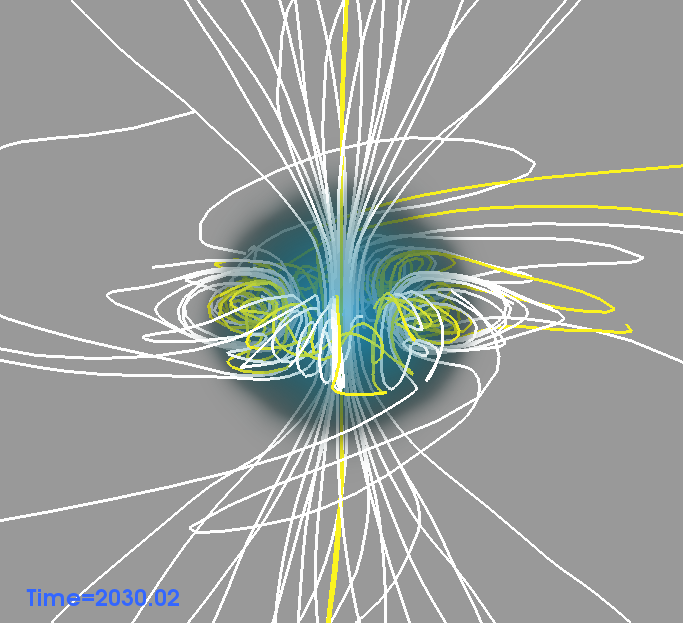}
		\caption{$t=10~{\rm ms}$}
	\end{subfigure}
	\hfill
	\begin{subfigure}{0.32\textwidth}
		\includegraphics[width=\textwidth]{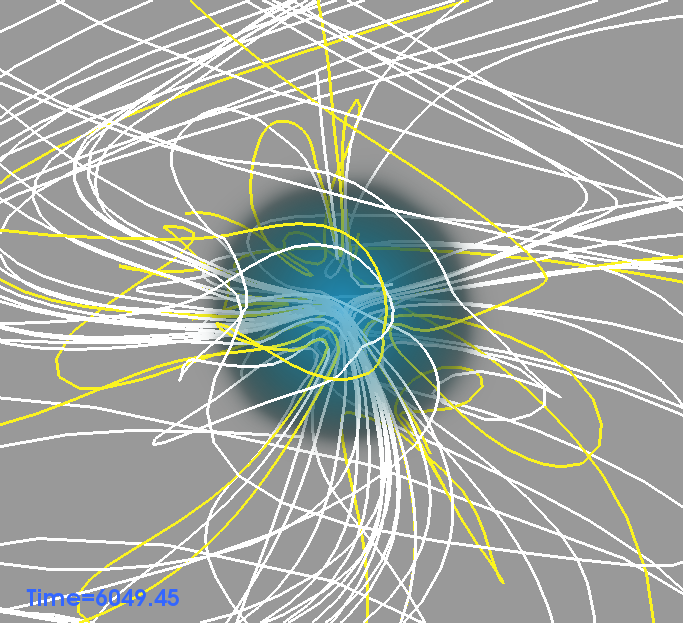}
		\caption{$t=30~{\rm ms}$}
	\end{subfigure}
	\hfill
	\begin{subfigure}{0.32\textwidth}
		\includegraphics[width=\textwidth]{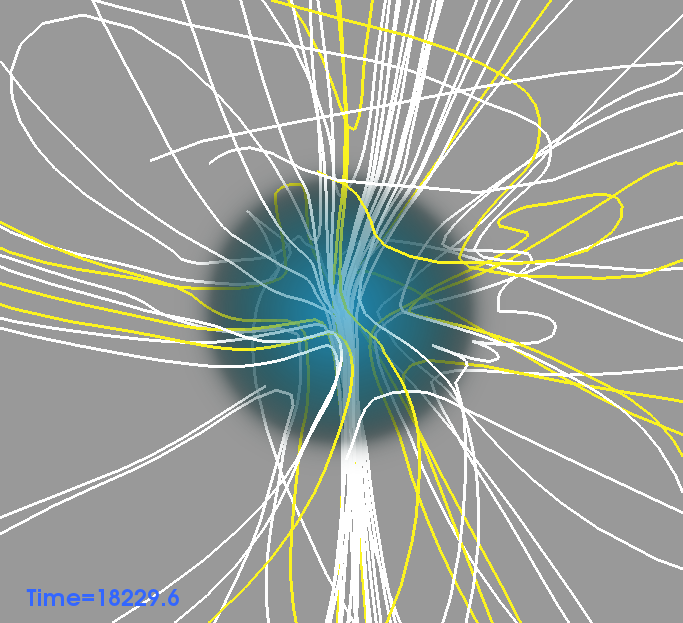}
		\caption{$t=90~{\rm ms}$}
	\end{subfigure}

	\caption{
		Three-dimensional renderings of model A2 with resistivity $\eta=10^{-6}$ (highly conducting) at 9 different instances of time. 
		The time labels in the plots are in the unit of $c = G = M_{\odot} = 1$. 
		White lines show the poloidal field lines while yellow lines show mixed (i.e. poloidal and toroidal) ones. 
		\label{fig:B_res6_evolution}
		}
\end{figure*}

\subsection{\label{sec:instability}Magnetic field evolutions and instability}

According to the Kruskal-Shafranov (KS) criterion (e.g., \cite{1978mit..book.....B}), cylindrical MHD configurations dominated by a toroidal magnetic field are highly unstable to the $m=1$ ``kink'' instability.
The KS criterion for instability is given by:
\begin{equation}\label{eq:ks}
	\left| \frac{B_{\rm tor}}{B_{\rm pol}} \right| \frac{r}{R} > 1,
\end{equation}
where $B_{\rm tor}$ and $B_{\rm pol}$ represent the toroidal and poloidal magnetic field strengths, $r$ and $R$ are the minor and major radii of the toroidal-dominated neutron star region (just below its surface), respectively.
Inside this toroidal-dominated region (which has a minor radius $r \approx 3$ and a major radius $R \approx 10.5$), the KS criterion is typically greater than 1, with a maximum value of approximately 1.7 at the center of the torus.
This suggests that the torus is initially unstable against the ``kink'' instability.
It is important to note that the criterion in equation~\eqref{eq:ks} is derived under assumptions of density, geometry, and Newtonian physics, which may not be applicable in the case of neutron stars.

\begin{figure*}
	\centering
	\begin{subfigure}{0.32\textwidth}
		\includegraphics[width=\textwidth]{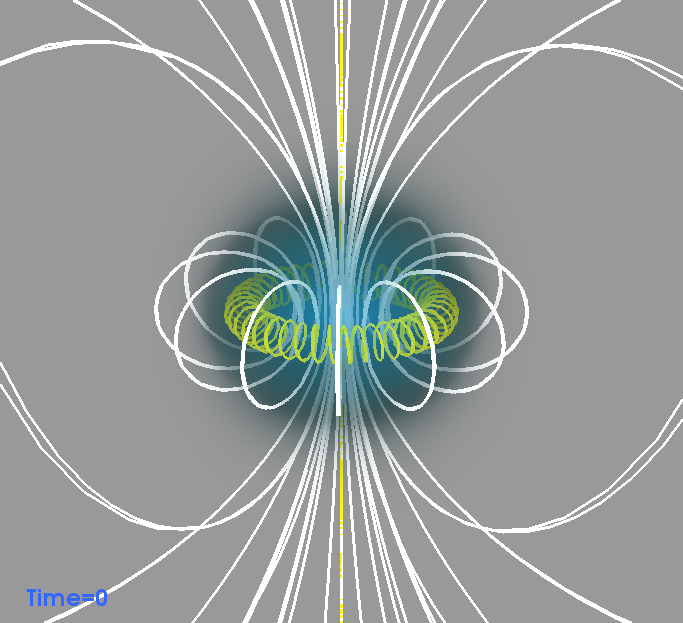}
		\caption{$t=0~{\rm ms}$}
	\end{subfigure}
	\hfill
	\begin{subfigure}{0.32\textwidth}
		\includegraphics[width=\textwidth]{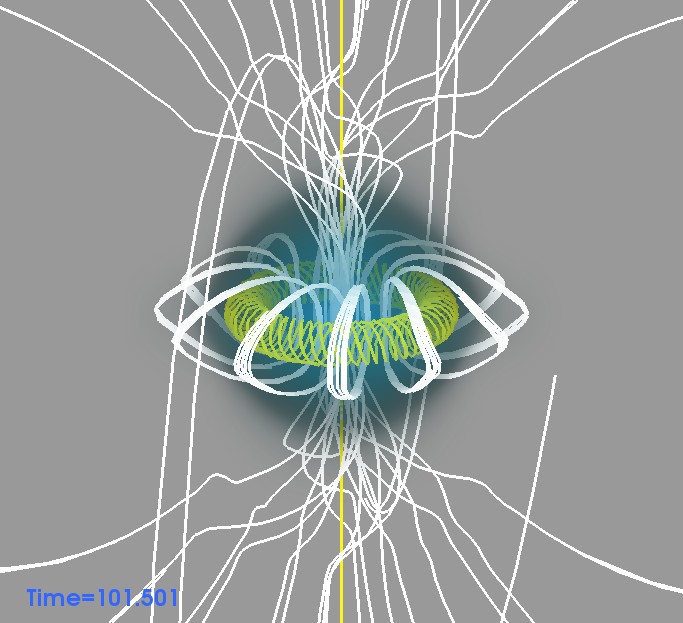}
		\caption{$t=0.5~{\rm ms}$}
	\end{subfigure}
	\hfill
	\begin{subfigure}{0.32\textwidth}
		\includegraphics[width=\textwidth]{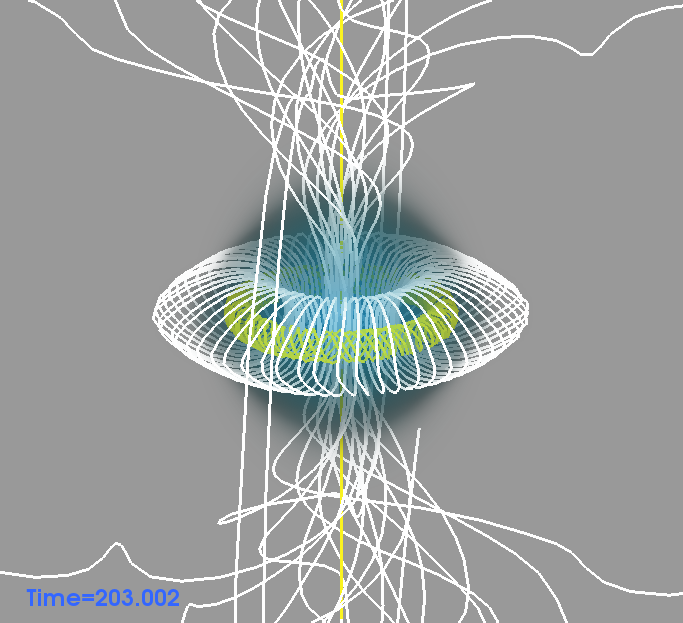}
		\caption{$t=1.0~{\rm ms}$}
	\end{subfigure}

	\begin{subfigure}{0.32\textwidth}
		\includegraphics[width=\textwidth]{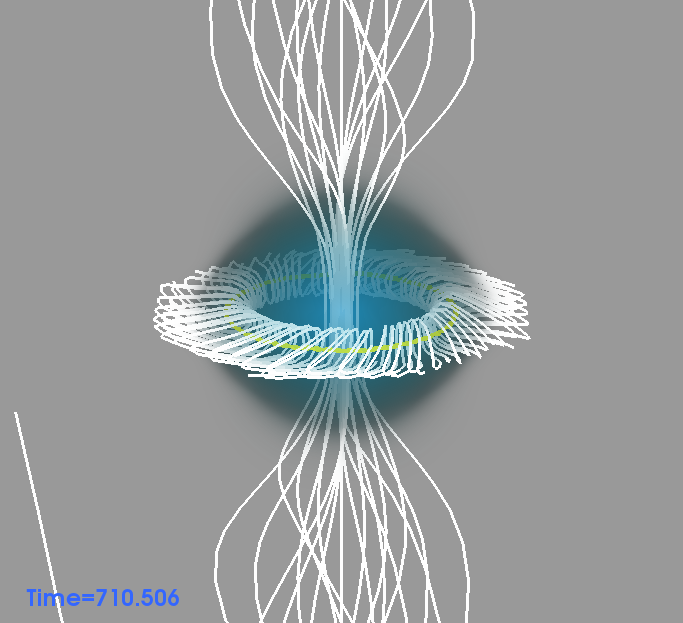}
		\caption{$t=3.5~{\rm ms}$}
	\end{subfigure}
	\hfill
	\begin{subfigure}{0.32\textwidth}
		\includegraphics[width=\textwidth]{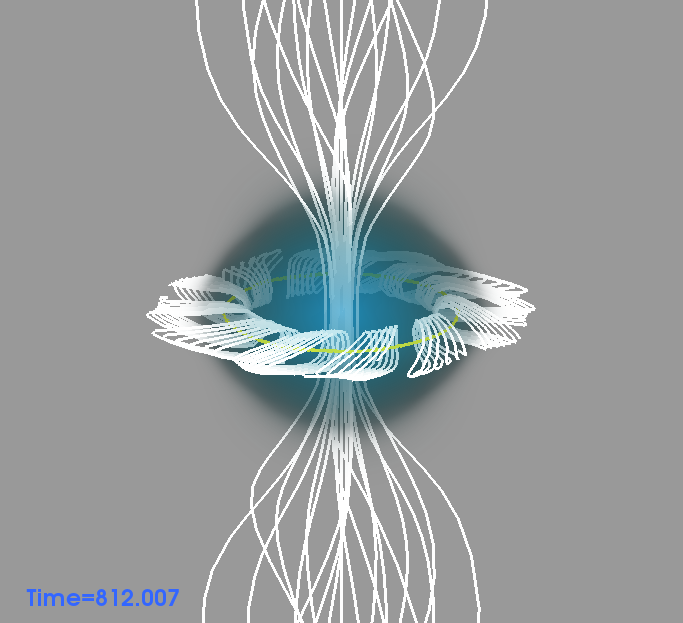}
		\caption{$t=4~{\rm ms}$}
	\end{subfigure}
	\hfill
	\begin{subfigure}{0.32\textwidth}
		\includegraphics[width=\textwidth]{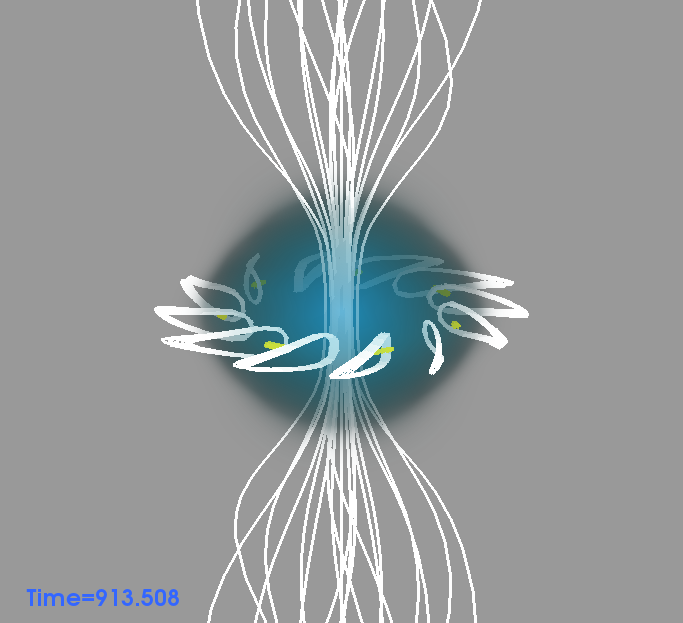}
		\caption{$t=4.5~{\rm ms}$}
	\end{subfigure}

	\begin{subfigure}{0.32\textwidth}
		\includegraphics[width=\textwidth]{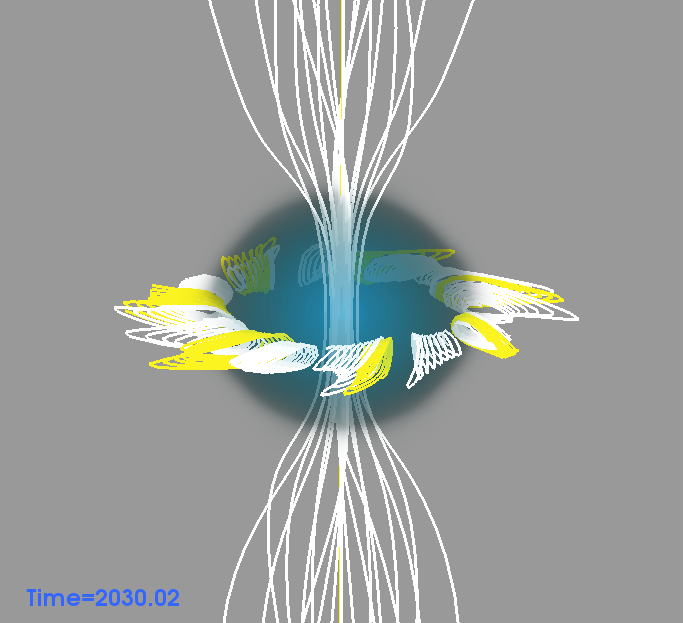}
		\caption{$t=10~{\rm ms}$}
	\end{subfigure}
	\hfill
	\begin{subfigure}{0.32\textwidth}
		\includegraphics[width=\textwidth]{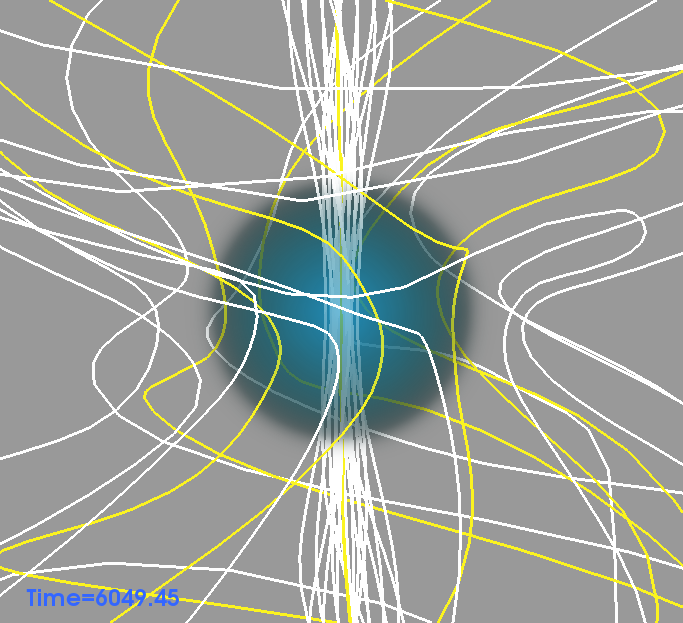}
		\caption{$t=30~{\rm ms}$}
	\end{subfigure}
	\hfill
	\begin{subfigure}{0.32\textwidth}
		\includegraphics[width=\textwidth]{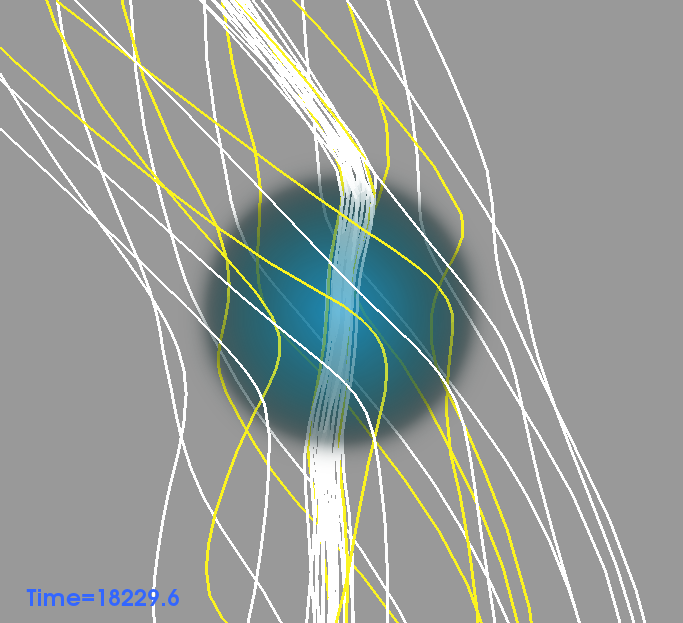}
		\caption{$t=90~{\rm ms}$}
	\end{subfigure}

	\caption{
		Similar plots as Fig.~\ref{fig:B_res6_evolution} but with the resistivity $\eta=10^{-2}$ (highly resistive) case. 
		\label{fig:B_res2_evolution}
		}
\end{figure*}

\begin{figure*}
	\centering
	\begin{tabular}{@{}lccc@{}}
		{} & $t=5~{\rm ms}$ & $t=20~{\rm ms}$ & $t=30~{\rm ms}$  \\
		\rotatebox{90}{$\eta = 10^{-6}$} &
			\includegraphics[width=0.32\textwidth]{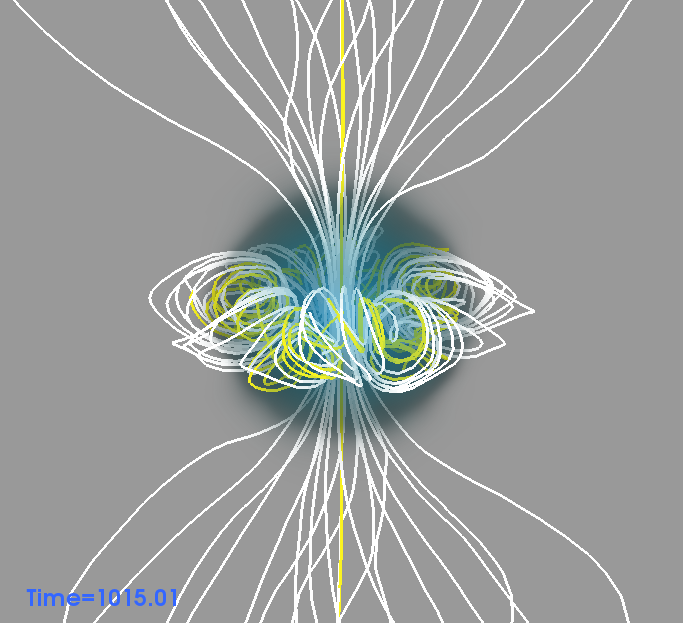} &
			\includegraphics[width=0.32\textwidth]{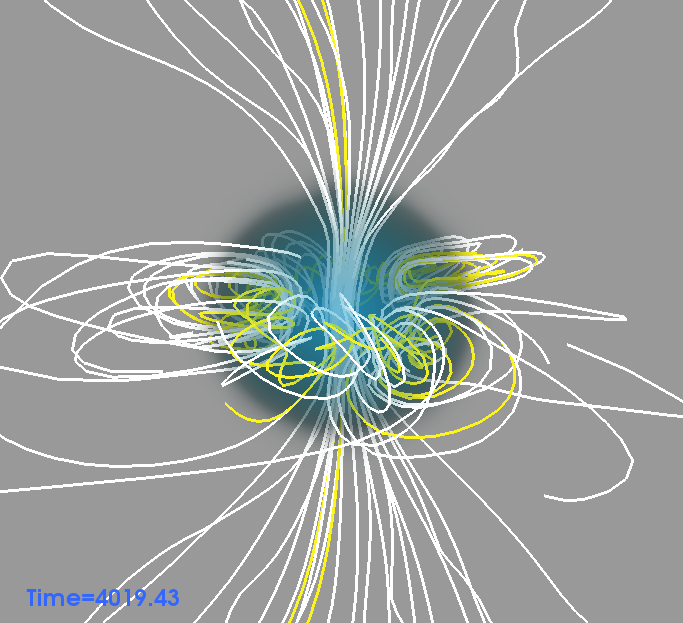} &
			\includegraphics[width=0.32\textwidth]{20240303_res_1e6_B2_l4_3d_683x623_0300.png} \\
		\rotatebox{90}{$\eta = 10^{-4}$} &
			\includegraphics[width=0.32\textwidth]{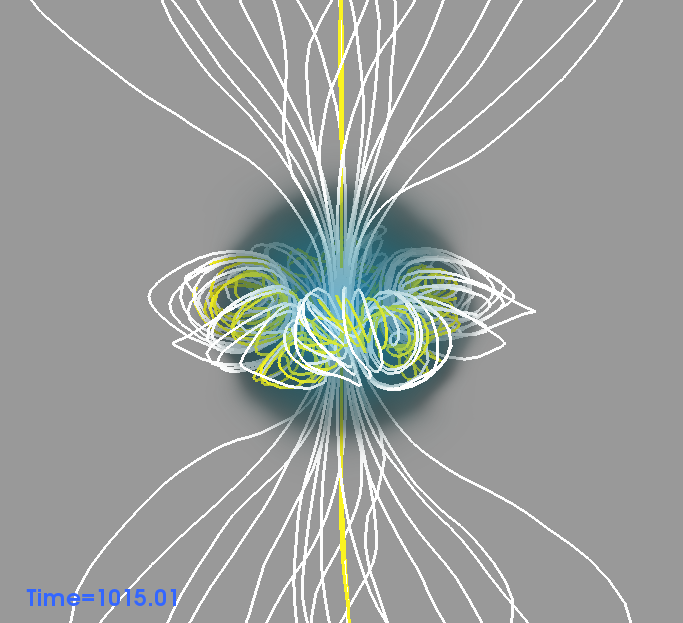} &
			\includegraphics[width=0.32\textwidth]{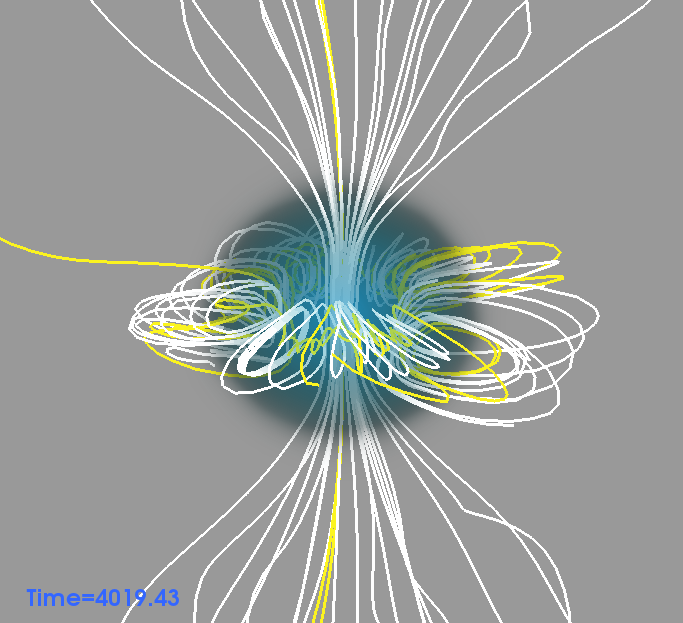} &
			\includegraphics[width=0.32\textwidth]{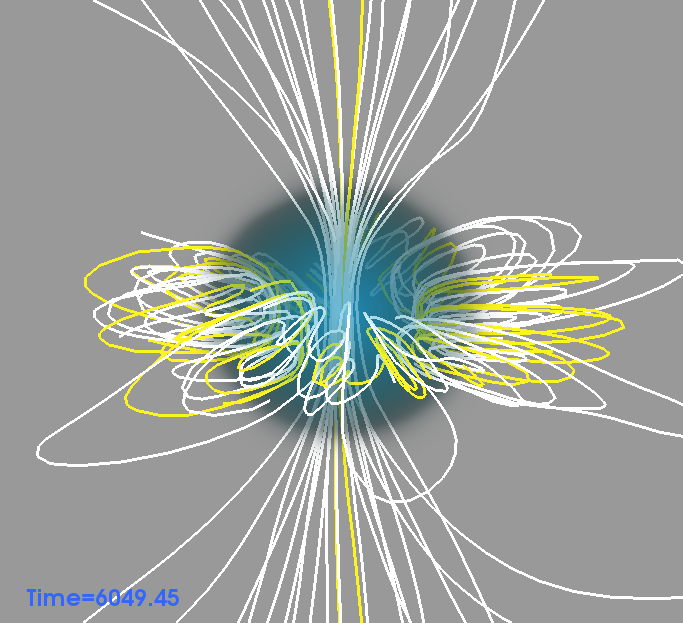} \\
		\rotatebox{90}{$\eta = 10^{-3}$} &
			\includegraphics[width=0.32\textwidth]{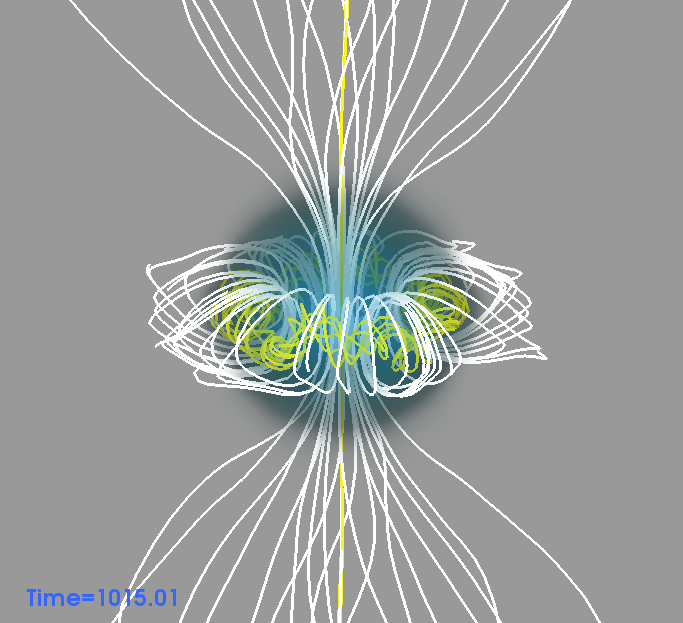} &
			\includegraphics[width=0.32\textwidth]{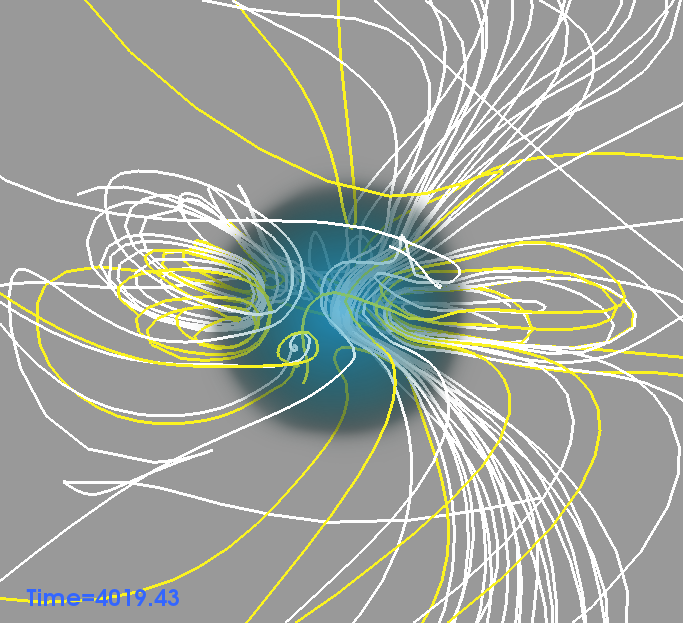} &
			\includegraphics[width=0.32\textwidth]{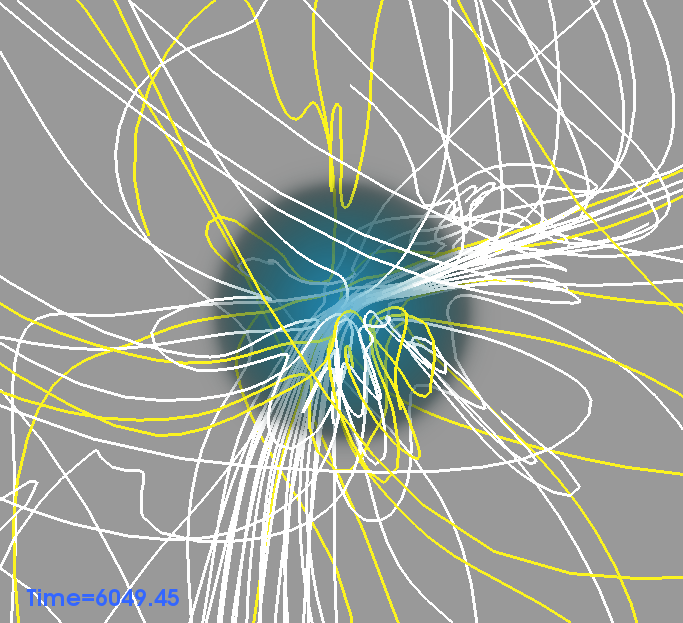} \\
		\rotatebox{90}{$\eta = 10^{-2}$} &
			\includegraphics[width=0.32\textwidth]{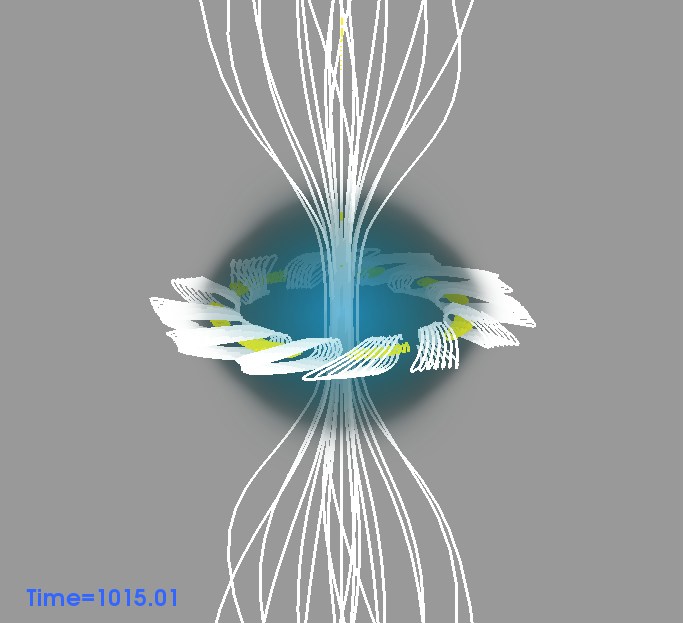} &
			\includegraphics[width=0.32\textwidth]{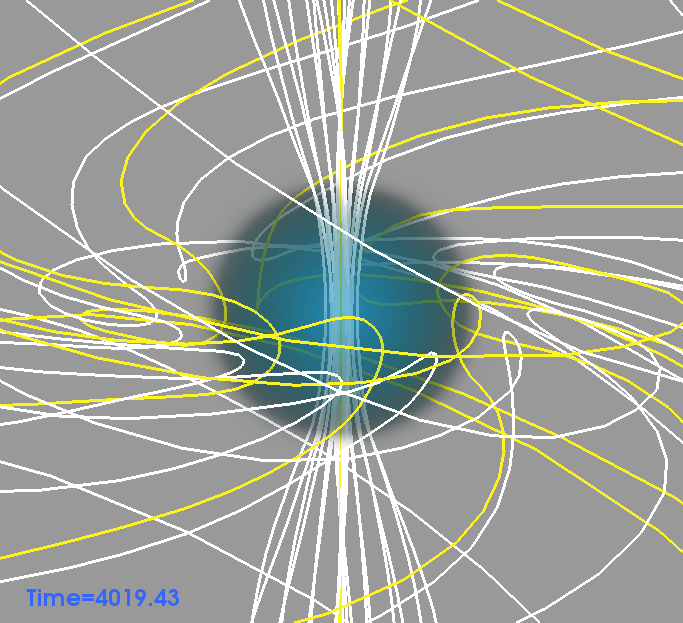} &
			\includegraphics[width=0.32\textwidth]{20240303_res_1e2_B2_l4_3d_683x623_0300.png} \\
	\end{tabular}
	\caption{
A comparison between evolutions with different resistivity $\eta = 10^{-6}, 10^{-4}, 10^{-3}, 10^{-2}$ shown from top to bottom rows, at $t=5,\,20,\,30~\rm{ms}$ shown from left to right columns.
		\label{fig:B_evolution}
	  }
\end{figure*}

To demonstrate how different instabilities grow at different time, in Fig.~\ref{fig:B_res6_evolution}, we show the three-dimensional renderings of the star with resistivity $\eta=10^{-6}$ (highly conducting case) at different times as an example.
The $m=0$ instability (also known as ``sausage'' instability) is developed at the beginning of the evolution.
As shown at the top row in Fig.~\ref{fig:B_res6_evolution} (from 0~ms to 1~ms), the cross-section of the toroidal magnetic field which is confined inside the star (the yellow ``spring'' surrounding the rotational axis) is changing in every direction, like breathing. 
The $m=1$ (``kink'') instability, is developed afterwards.
As shown at the middle row (from 3.5~ms to 4.5~ms), instead of surrounding the rotational axis strictly on the $z=0$ plane, the yellow ``spring'' oscillates around the $z=0$ plane.
At the same time, the poloidal closed white loops at the star surface starts to be twisted together with the internal fields. 
Despite the instabilities, the overall poloidal-like structure still remains up to 30~ms, and is destroyed completely around 50~ms, as shown at the bottom row.
Given the dynamical timescale by which these instabilities develop, the decay of the electromagnetic, kinetic, and rotational energies are shown in Figs.~\ref{fig:all_energy}, \ref{fig:rhomax}, and \ref{fig:beta_and_Eem_over_T}.
In contrast, the ``sausage'' and ``kink'' instabilities mentioned earlier do not develop in the highly resistive case with $\eta=10^{-2}$, as shown in Fig.~\ref{fig:B_res2_evolution}.
Both the closed magnetic field lines inside and outside the star are flatten onto $xy$-plane.
At the end of the simulation, the overall poloidal-like structure still remains, and the magnetic fields concentrate at the pole of the star. 
Although it is expected that eventually a stable equilibrium will establish itself by means of Taylor relaxation, we did not reach that point. 
A comparison between the different resistivity evolutions at specific instances is shown in Fig.~\ref{fig:B_evolution}.
  
As described in Sec.~\ref{sec:sim}, the region surrounding the star is initially filled with a low-density gas with a constant magnetic-to-gas pressure ratio.
Consequently, the evolution of the magnetic field in these low-density regions depends on the properties of the gas, such as the rest-mass density and the magnetic-to-gas pressure ratio.
The impact of parameter choices, as well as the development of more realistic treatments, which are crucial for neutron star modeling, will be addressed in future work.

\begin{figure*}
	\centering
	\includegraphics[width=\textwidth, angle=0]{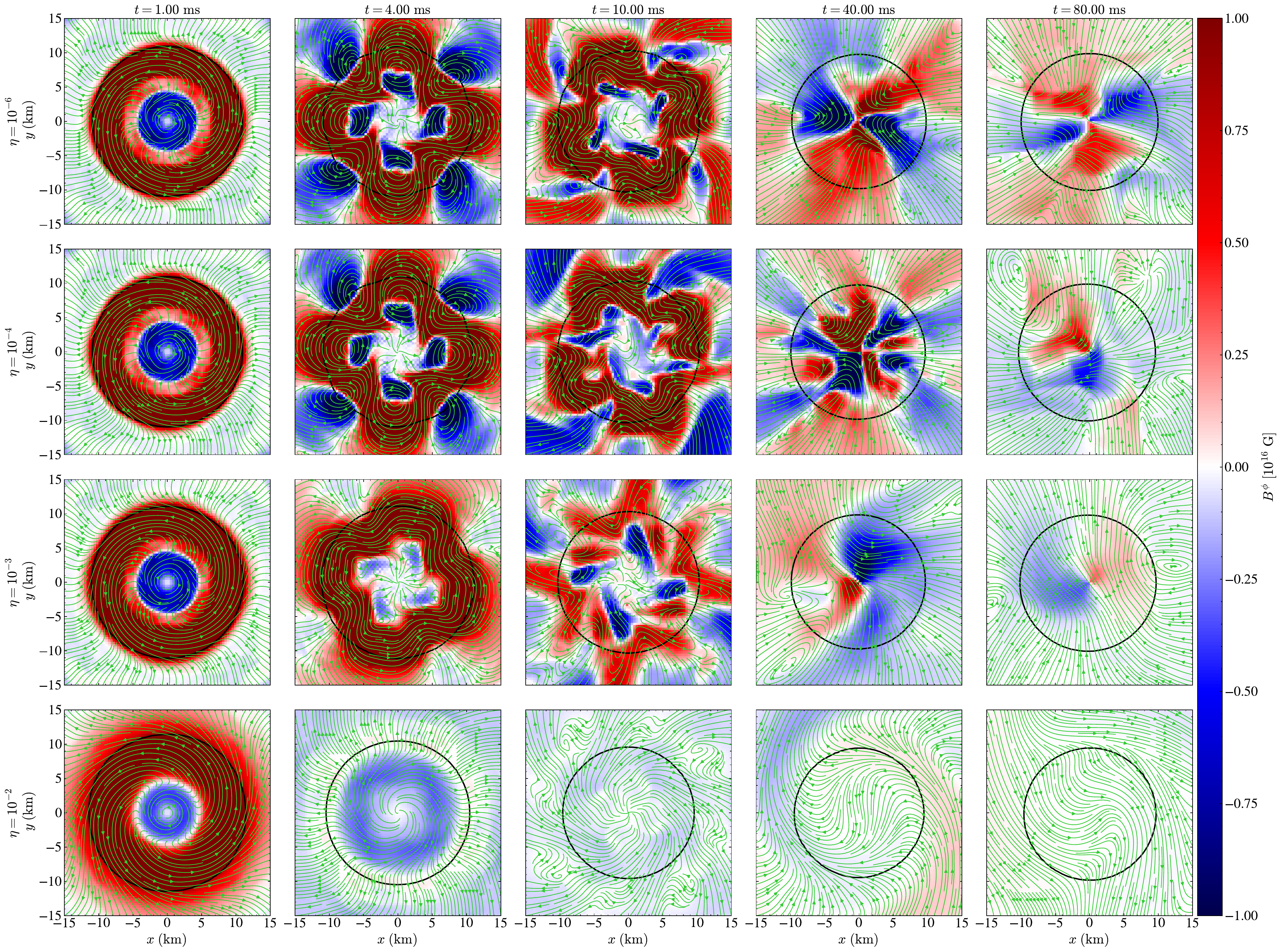}
	\caption{
		Toroidal magnetic field strength $B^{\phi}$ (colours) and the magnetic field lines (streamlines) on the $xy$-plane at different time with different resistivity $\eta$.
		The black solid lines show the surface of the neutron star.
		Cases with different resistivity $\eta = 10^{-6}, 10^{-4}, 10^{-3}, 10^{-2}$ are shown from top to bottom rows, while the snapshots at $t=1, 4, 10, 40, 80~\rm{ms}$ are shown from left to right columns.
		\label{fig:Bphi_xy_time_series}
		}
\end{figure*}

\begin{figure*}
	\centering
	\includegraphics[width=\textwidth, angle=0]{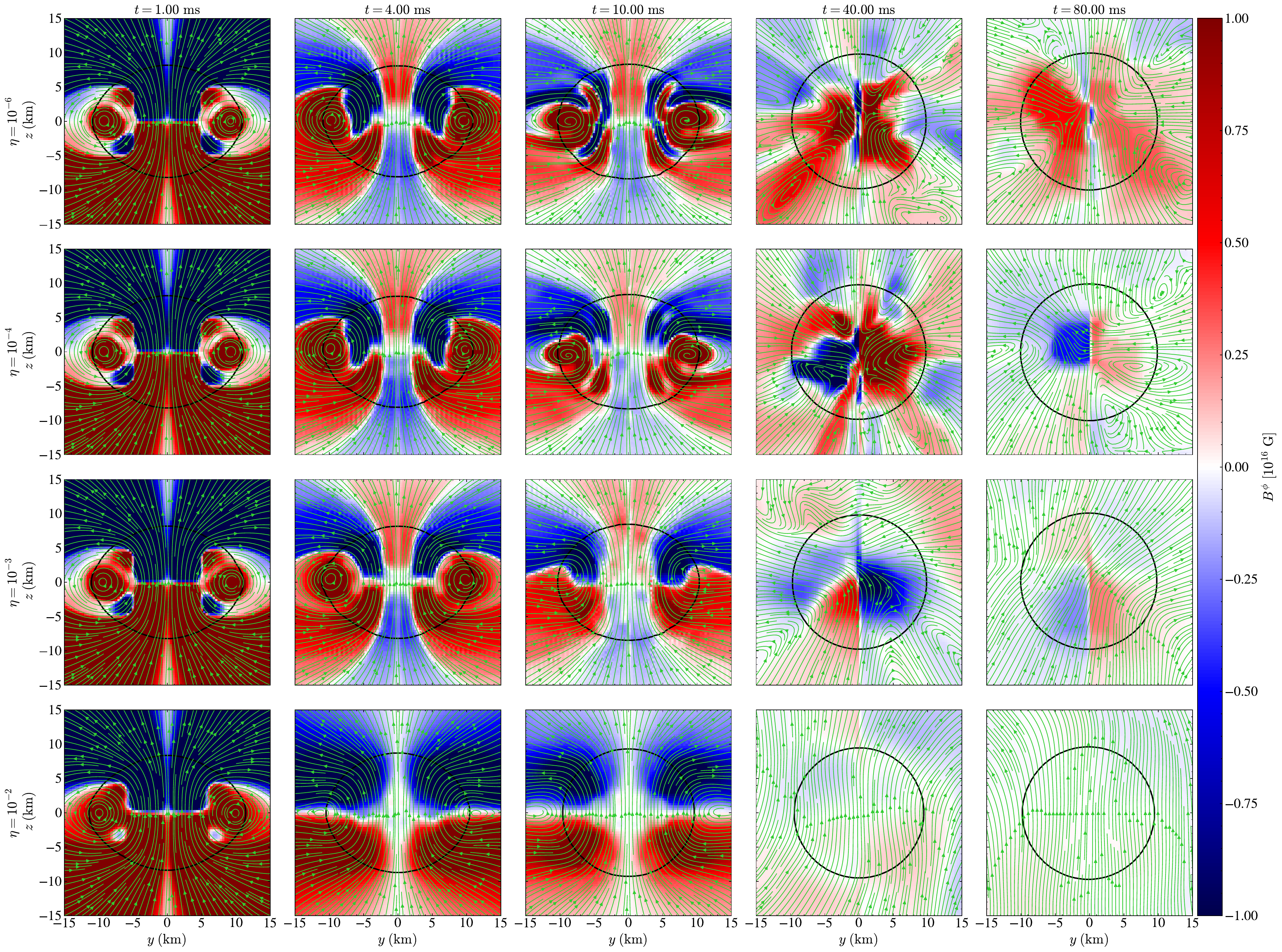}
	\caption{
		Similar plots as Fig.~\ref{fig:Bphi_xy_time_series} but on the $yz$-plane.
		The colours show the toroidal magnetic field strength $B^{\phi}$, while the streamlines show the magnetic field lines. 
		The black solid lines show the surface of the neutron star.
		As discussed in Fig.~\ref{fig:Bphi_xy_time_series}, as the instability grows, it destroys the magnetic field symmetries, and develops multipolar topologies.
		\label{fig:Bphi_yz_time_series}
		}
\end{figure*}

Figs.~\ref{fig:Bphi_xy_time_series} and \ref{fig:Bphi_yz_time_series} compare the toroidal magnetic field strength $B^{\phi}$ and the magnetic field lines on the $xy$- and $yz$-planes at different instances and different resistivities $\eta$.
In the highly conducting case ($\eta = 10^{-6}$), the ``sausage'' and ``kink'' instabilities develop very quickly soon after the simulation is lunched. 
At early time $t \sim 4~\rm{ms}$, the toroidal component creates vertex-like structures as reported in the literature (e.g. \cite{2011ApJ...736L...6C, 2012ApJ...760....1C, 2020MNRAS.495.1360S}).
At late times, the symmetries of the magnetic fields are mostly destroyed. Since the existence of resistivity dissipates the strength of the magnetic fields, it affects how the instability is developed. 
The existence of resistivity dissipates the strength of the magnetic fields, and delays the growth of instabilities.
In the highly resistive cases ($\eta = 10^{-2}$), at the late time ($t=80~\rm{ms}$) the field lines are aligned and uniformly distributed. 
Note that, the oblateness of the star sensitively depends on the strength of the poloidal magnetic fields and the rotations. 
As the star losses the magnetic pressure and spins down (e.g. see Fig.~\ref{fig:combined_omega}), it becomes less oblate and asymptotically become quasi-spherically symmetric.
  

In addition to the magnetic instabilities,  the dynamical bar mode instability will occur when the rotational kinetic-to-gravitational-energy ratio $T_{\rm rot} / \lvert \mathcal{W} \rvert$ of a rotating neutron star is larger than a certain threshold.
In general, the threshold is found to be around 0.25 to 0.1 depending on the differential rotation law \cite{2000ApJ...542..453S, 2007PhRvD..75d4023B, 2007CQGra..24S.139O}, although non-axisymmetric instabilities for values as low as 0.01 have also been found \cite{Shibata:2002mr,Shibata:2003yj}. 
These so-called shear instabilities depend on $T_{\rm rot} / \lvert \mathcal{W} \rvert$ and the degree of differential rotation \cite{Watts2005}. 
In the top panel of the Fig.~\ref{fig:beta_and_Eem_over_T}, we plot the rotational kinetic-to-gravitational-energy ratio $T_{\rm rot} / \lvert \mathcal{W} \rvert$ of our star for all resistivities. 
This ratio starts below 0.1, and drops several orders or magnitude, thus we do not observe any development of the bar mode instability. 
This can also be seen in Figs.~\ref{fig:B_res6_evolution} and \ref{fig:B_res2_evolution} where the stars evolve towards spherical symmetry.

It has been shown that Tayler instability can be triggered in \emph{toroidal dominated} magnetized neutron stars \cite{2008PhRvD..78b4029K, 2011A&A...532A..30K}.
In the axisymmetric study \cite{2008PhRvD..78b4029K}, the authors show that the axisymmetric Tayler instability is triggered when the ratio $E_{\rm EM} / T_{\rm rot} \gtrsim 0.2$.
The model considered in this work is poloidal dominated, and as shown in the bottom panel in Fig.~\ref{fig:beta_and_Eem_over_T}, the ratio $E_{\rm EM} / T_{\rm rot}$ is beyond 0.2 initially,
increasing as the system evolves.
All our simulations are unstable against the Tayler instability, pointing to the fact that a toroidal dominated magnetic field is not a necessary condition for its development. 
Also, despite the fact that the star is rotating sufficiently fast, the instability is not suppressed.
As shown by Frieman and Rotenberg \cite{Frieman1960}, rigid-body rotation has a significant effect on hydromagnetic equilibria when the fluid-flow velocity is of the same order as the Alfv\'en velocity. 
In our model the initial maximum fluid velocity is $0.284c$, while the initial Alfv\'en velocity is $0.08 c$, i.e. an order of magnitude smaller.
Therefore, it is not expected that rotation can stabilize the developed instabilities, consistent with our simulations.

\begin{figure*}
	\centering
	\includegraphics[width=\textwidth, angle=0]{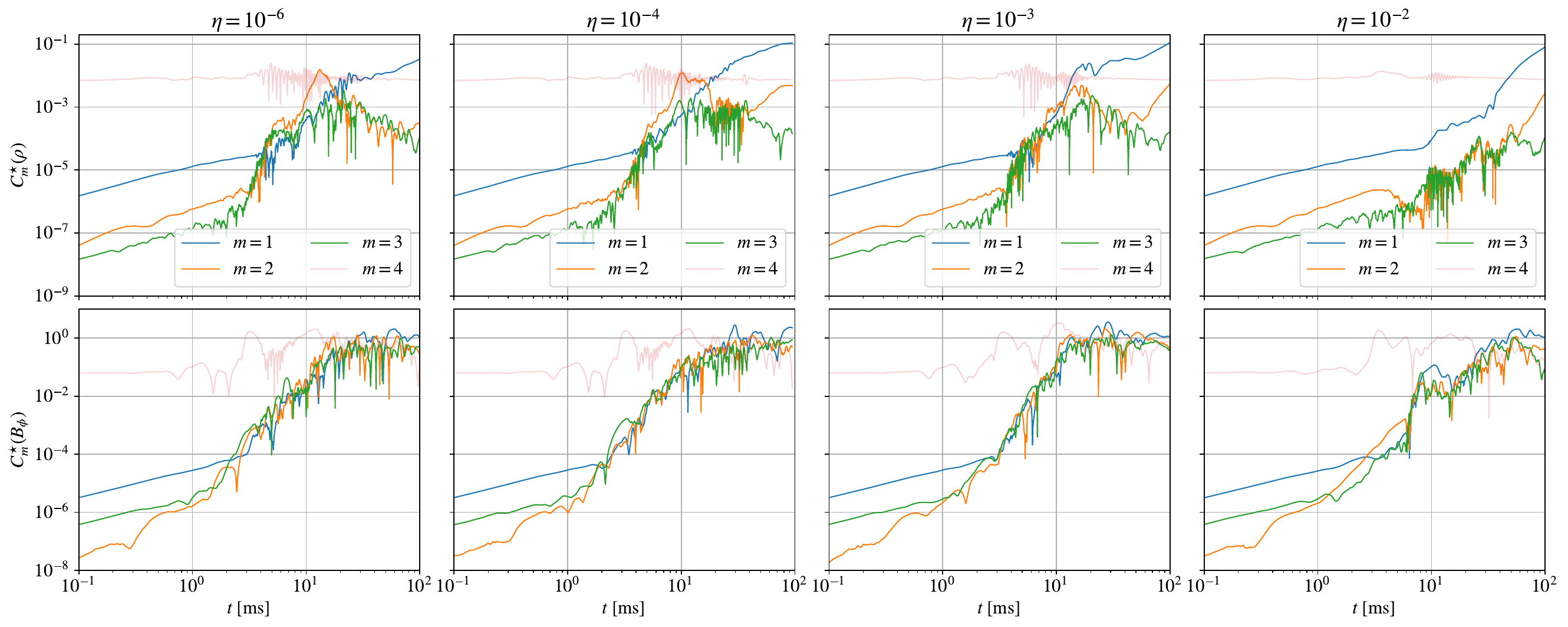}
	\caption{
		Volume-integrated azimuthal mode decomposition of the conserved rest-mass density $W \rho$ (upper panels) and toroidal magnetic field $B^{\phi}$ (lower panels).
		Normalized modes with azimuthal number in the range $m=1,2,3,4$ are shown.
		Cases with different resistivity $\eta = 10^{-6}, 10^{-4}, 10^{-3}, 10^{-2}$ are shown from left to right columns.
		\label{fig:cm}
		}
\end{figure*}

To better understand the development and saturation of the instability, we compute the volume-integrated azimuthal mode decomposition of rest mass density $D:= W \rho$ and toroidal magnetic field $B^{\phi}$.
The normalized modes for both conserved rest-mass density and toroidal magnetic field with azimuthal number in the range $m=1,2,3,4$ with different resistivity are shown in Fig.~\ref{fig:cm}. 
Note that the $m=4$-modes are dominated by the Cartesian grid induced perturbation (see e.g. \cite{2007PhRvL..98z1101O, 2020PhRvD.102d4040X, 2023PhRvD.108j4005J}), which remains mostly at the same level throughout the simulations.

\begin{figure*}
	\centering
	\includegraphics[width=\textwidth, angle=0]{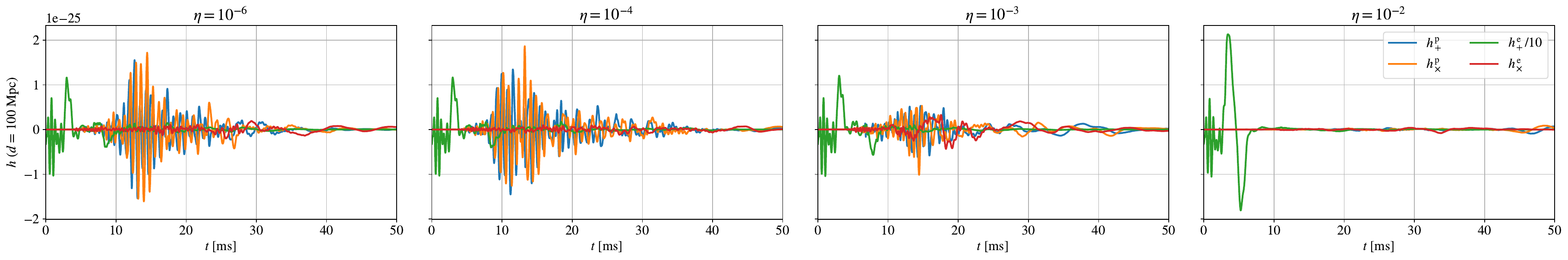}
	\caption{
		Time evolution of the gravitational wave amplitude $h$ observed at polar axis and equatorial plane with different polarizations of the strongly magnetized rapidly rotating neutron star model in Table~\ref{tab:table_ID} with different resistivities $\eta$.
		Since the magnetar system is very axisymmetric, the gravitational wave strain $h^{\rm e}_+$ is expected to be much larger than others.
		Therefore, $h^{\rm e}_+$ is divided by 10 for better visualisation.
		As the $m=2$-mode of rest mass density expected to be the dominating mode of gravitational waves emissions, the suppression of the $m=2$-mode shown in Fig.~\ref{fig:cm} should also be reflected in the gravitational wave amplitude.
		As expected, the existence of resistivity suppresses the gravitational wave amplitude, as a direct consequence of the suppression of the $m=2$-mode.
		\label{fig:gw}
		}
\end{figure*}

The upper panel of Fig.~\ref{fig:cm} compares the volume-integrated azimuthal mode decomposition of the conserved rest-mass density $W \rho$ with different resistivity $\eta$.
In all cases, the~$m=1$-mode~grows at a similar rate and does not saturate.
This implies that the one-arm spiral instability~\citep{2005ApJ...625L.119O, 2006ApJ...651.1068O, 2017ApJ...840...16K} grows throughout the entire simulations, and is insensitive to the choice of resistivity.
On the other hand, the $m=2,3$-modes reach to their peak values at around 10~ms except for the highest resistive case $\eta = 10^{-2}$.
The existence of resistivity suppress the peak value of $m=2$-mode, and boosting its growth at later times. 
Since the $m=2$-mode is usually the dominating mode of gravitational wave emission, this suppression can also be seen in the gravitational waves signals, as shown in Fig.~\ref{fig:gw}.
Unlike the conserved rest-mass density modes, the mode decomposition of the toroidal magnetic field $B^{\phi}$ behaves similarly for all resistivities.
As shown in the lower panel of Fig.~\ref{fig:cm}, the $m=1,2,3$-modes grow at a similar rate, and saturate around 10~ms.

\begin{figure}
	\centering
	\includegraphics[width=\columnwidth, angle=0]{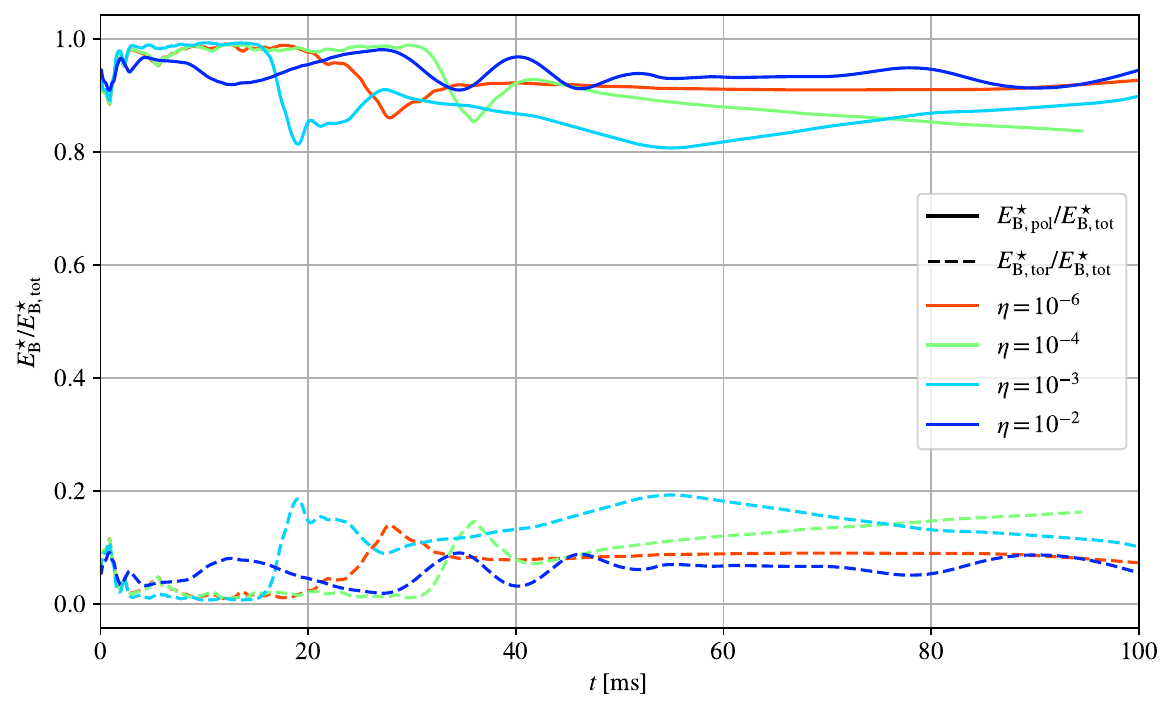}
	\caption{
		Time evolution of the magnetic energy ratios $E^{\star}_{B, \rm{pol}}/E^{\star}_{B, \rm{tot}}$ (solid lines) and $E^{\star}_{B, \rm{tor}}/E^{\star}_{B, \rm{tot}}$ (dashed lines) of the star with different resistivities $\eta$. 
		Despite the varying resistivity and resulting differences in instability development across these models, the energy ratios of the poloidal to toroidal fields consistently remain around $9:1$ throughout the evolution.	
		\label{fig:eb_ratios}
		}
\end{figure}

\section{\label{sec:conclusions}Conclusions}
We reported long-term general-relativistic resistive-MHD simulations of self-consistent rotating neutron stars with ultra-strong mixed poloidal and toroidal fields.
One of the major effects due to resistivity is the Ohmic dissipations of magnetic fields. 
Since the magnetic field evolution sensitively depends on the development of various instabilities, which in turn depend on the strength of the magnetic field, the existence of non-zero resistivity is expected to alter the entire evolution significantly.
To explore the long-term effects due to resistivity, we performed dynamical simulations of a magnetized neutron star up to 100~ms with 4 different resistivities, and study the evolution of magnetic fields, instabilities, and field geometry.

We found that the resistivity of the star can significantly alter the development of the magnetohydrodynamical instabilities.
In particular, we showed that the dissipation of the magnetic energy is dominated by the choice of resistivity. 
As shown in Fig.~\ref{fig:rhomax}, the higher the resistivity, the larger the magnetic energy dissipation.
A direct consequence of reducing the strength of a magnetic field is the increase of the Alfv{\'e}n timescale, and hence the delay of the growth of the instability.
Figs.~\ref{fig:cm} and \ref{fig:gw} demonstrate the suppression of the instability, indicating that gravitational wave emission is positively correlated with resistivity.

We also found that the magnetic field geometry evolution sensitively depends on the resistivity.
The magnetic field evolution depends critically on the growth of instability, which is shown to be related to resistivity.
For instance, in the highly conducting cases, instability develops and break the symmetries in a short timescale, resulting in a complicated multipolar field structure.
In contrast, in the highly resistive cases, the instability is suppressed, the field lines are relatively ordered.
At the end of the simulation, a large scale uniform dipolar structure is formed.
Surprisingly, although the evolution of the magnetic field geometry is very different in all cases, the poloidal-to-toroidal field energy ratio remains quantitatively $9:1$ throughout the simulations, as shown in Fig.~\ref{fig:eb_ratios}.

We note here that the magnitude of the magnetic field in our initial models is unrealistically large.
The reason for adopting such large magnetic fields is to make the Alfv\'en timescale sufficiently small so  that our evolutions reach a meaningful point within the finite amount of computing resources.
On the other hand, given the fact that the stability of extremely magnetized neutron stars is currently unknown, by choosing a large magnetic field we risk evolving objects that are magnetically unstable.
In the future we will perform longer simulations with smaller (and more realistic) magnetic fields.

We will further investigate the impact on neutron star evolution due to resistivity.
For example, the resistivity is expected to be very low at the inner part of the neutron star, while it can be significantly higher at the surface or the region surrounding the star \citep{2019LRCA....5....3P}.
In a more realistic consideration, resistivity should be at least a function of rest mass density.
Moreover, exploration of the parameter spaces of neutron stars (e.g. rotations, initial field strength and configurations) is necessary to complete the picture.
Finally, high resolution simulations are also needed to better capture the growth of instability.
These will be left as our future work.

\begin{acknowledgments}
P.C.-K.C. gratefully acknowledges support from National Science Foundation (NSF) Grant PHY-2020275 (Network for Neutrinos, Nuclear Astrophysics, and Symmetries (N3AS)).
A.T. is supported in part by NSF Grants No. PHY-2308242 and No. OAC-2310548 to the University of Illinois Urbana-Champaign. 
A.T. acknowledges support from the National Center for Supercomputing Applications (NCSA) at the University of Illinois Urbana-Champaign through the NCSA Fellows program.
M.R. acknowledges support by the Generalitat Valenciana Grant CIDEGENT/2021/046 and CIGRIS/2022/126 and by the Spanish Agencia Estatal de Investigaci\'on (Grant PID2021-125485NB-C21). 
J.C.L.C acknowledges support from the Villum Investigator program supported by the VILLUM Foundation (grant no. VIL37766) and the DNRF Chair program (grant no. DNRF162) by the Danish National Research Foundation.
K.U. is supported by JSPS Grant-in-Aid for Scientific Research (C) 22K03636 to the University of the Ryukyus.

The simulations in this work have been performed on the UNH supercomputer Marvin, also known as Plasma, which is supported by NSF Major Research Instrumentation (MRI) program under grant number AGS-1919310. 
This work also used Expanse cluster at San Diego Supercomputer Centre through allocation PHY240086 from the Advanced Cyber infrastructure Coordination Ecosystem: Services \& Support (ACCESS) program~\citep{10.1145/3569951.3597559}, which is supported by National Science Foundation grants 2138259, 2138286, 2138307, 2137603, and 2138296. 
Finally, we acknowledge computational resources and technical support of the Spanish Supercomputing Network through the use of MareNostrum at the Barcelona Supercomputing Center (AECT-2023-1-0006).

The data of the simulations were post-processed and visualised with 
\texttt{yt}~\citep{2011ApJS..192....9T},
\texttt{NumPy}~\citep{harris2020array}, 
\texttt{pandas}~\citep{reback2020pandas, mckinney-proc-scipy-2010},
\texttt{SciPy}~\citep{2020SciPy-NMeth},
\texttt{Matplotlib}~\citep{2007CSE.....9...90H, thomas_a_caswell_2023_7697899}, and
\texttt{VisIt}~\citep{Childs_High_Performance_Visualization--Enabling_2012}.
\end{acknowledgments}

\appendix
\section{\label{sec:code_tests}Comparison to fully general relativistic simulations}
As discussed in section~\ref{sec:methods}, the initial data we use in this work is fully general relativistic.
Given that the \texttt{Gmunu} code adopts the conformally flat approximation, it is necessary to verify how this approximation might affect the evolution. 
To this end, we compare simulations of non-conformally-flat initial data generated using the \texttt{IllinoisGRMHD} and \texttt{Gmunu} codes. 
The numerical setup for \texttt{Gmunu} is the same as described in Section~\ref{sec:methods}, except for the time integration method. 
Specifically, we use the explicit time-stepping scheme SSPRK3~\citep{1988JCoPh..77..439S} for pure hydrodynamical and ideal magnetohydrodynamical cases. 
On the other hand, the numerical setup for~\texttt{IllinoisGRMHD} is the same described in~\citep{2022PhRvL.128f1101T}.

\subsection{Non-magnetized evolution}
In this subsection, we compare the evolutions of a non-magnetized rapidly rotating neutron star with \texttt{Gmunu} and \texttt{IllinoisGRMHD} codes. 
The model employed in the case corresponds to a neutron stars with a central rest-mass density $\rho = 4.90\times 10^{14}~\rm g \cdot cm^{-3}$, a gravitational mass $M_0 = 1.83 M_\odot$, a radius $ R_{\rm NS} = 19.75~\rm km$ along the $x$ coordinate, and a rotational kinetic-to-gravitational-energy ratio of $7.15\times 10^{-2}$.
This equilibrium model is generated with a polytropic equation of state with $\Gamma = 2$ and $K=123.6$.
Resolution on the innermost level in both cases are about $\Delta x \approx 437.54~\rm{m}$.
In both cases, ideal-gas equation of state with $\Gamma = 2$ is used, and equatorial symmetry is adopted.

\begin{figure}
	\centering
	\includegraphics[width=\columnwidth, angle=0]{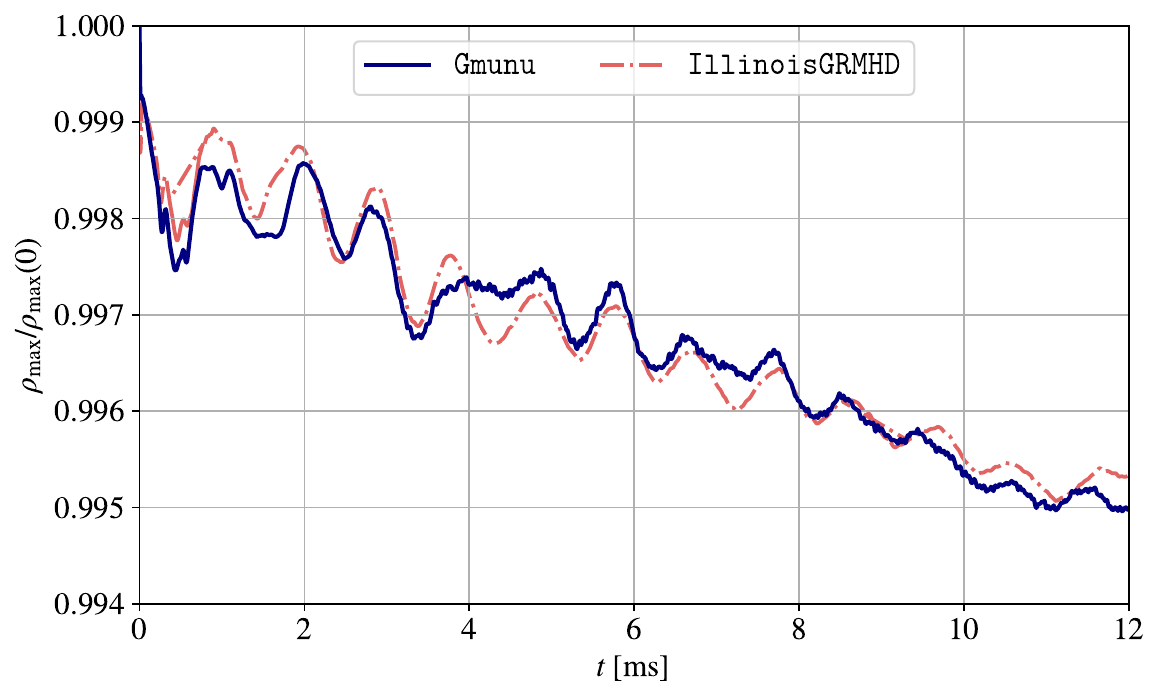}
	\caption{
		Comparison of the three-dimensional evolutions of a non-magnetized, rapidly rotating neutron star using the \texttt{Gmunu} and \texttt{IllinoisGRMHD} codes. 
		Here we compare the maximum rest-mass density normalized by its initial value, $\rho_{\max}/\rho_{\max}\left(0\right)$, from \texttt{Gmunu} (solid line) which adopts the conformally-flat approximation, and the \texttt{IllinoisGRMHD} (dashed line) which evolves in full general relativity.
		\label{fig:grhd_WL25_compare_rhomax}}
\end{figure}

Fig.~\ref{fig:grhd_WL25_compare_rhomax} compares the time evolutions of the neutron star using the \texttt{Gmunu} and \texttt{IllinoisGRMHD} codes. 
The simulations with \texttt{Gmunu} quantitatively agree with those from \texttt{IllinoisGRMHD}, despite the former adopting the conformally-flat approximation for evolution.
\begin{figure}
	\centering
	\includegraphics[width=\columnwidth, angle=0]{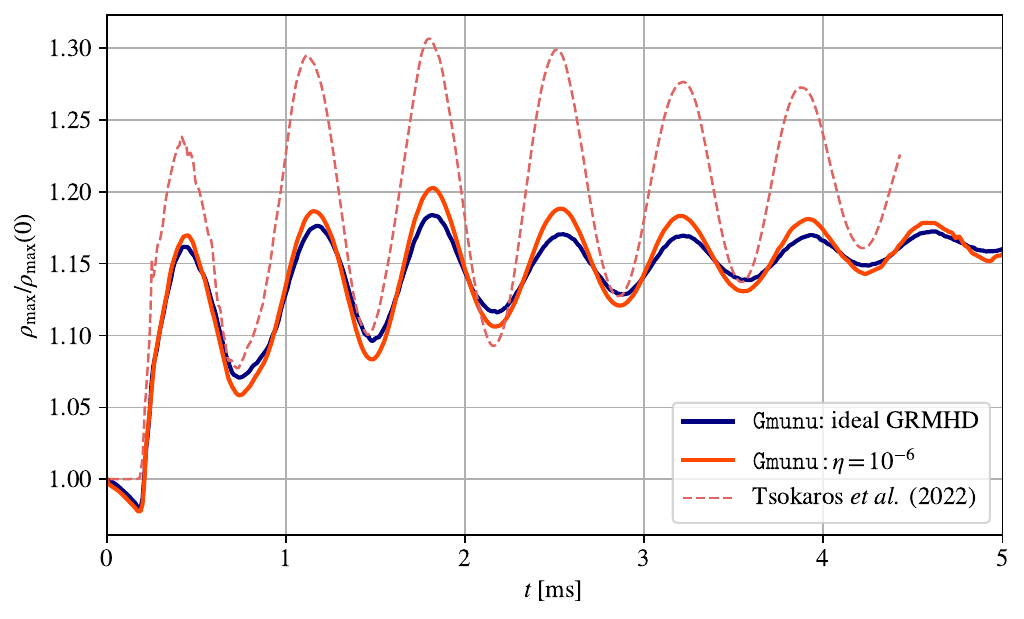}
	\caption{
		Comparison of the evolutions of a strongly magnetized, rapidly rotating neutron star using the \texttt{Gmunu} and \texttt{IllinoisGRMHD} codes. 
		This magnetar corresponds to the ``A2" model as described in~\cite{2022PhRvL.128f1101T}. 
		We compare the maximum rest mass density normalized by its initial value~$\rho_{\max}/\rho_{\max}\left(0\right)$. 
		The solid lines show the numerical results obtained by~\texttt{Gmunu}: the ideal MHD simulation is shown in navy, while the resistive simulation with low conductivity $\eta=10^{-6}$ is shown in red. 
		The dashed pink line represents the ideal magnetohydrodynamical evolution carried out using~\texttt{IllinoisGRMHD}, as presented in \cite{2022PhRvL.128f1101T}. 
		The results generated by \texttt{Gmunu} broadly agree with the reference solution produced by the ideal full GRMHD code \texttt{IllinoisGRMHD}.
		\label{fig:compare_A2_rhomax}
		}
\end{figure}

\subsection{Magnetized evolution}
In this subsection, we compare the evolutions of the same strongly magnetized rapidly rotating neutron star model ``A2'' with \texttt{Gmunu} and \texttt{IllinoisGRMHD} codes.
This neutron star is the ``A2'' model as described in section~\ref{sec:methods} and \cite{2022PhRvL.128f1101T}.
Note that, the finest grid size at the centre of the star in the case of \texttt{Gmunu} is $\Delta x \approx 346 ~\rm{m}$, while it is about 87~m in the case of \texttt{IllinoisGRMHD}~\cite{2022PhRvL.128f1101T}.
Fig.~\ref{fig:compare_A2_rhomax} compares the time evolutions of the maximum rest mass density with different codes, where the rest mass density is normalised by its initial value.

The solid lines in Fig.~\ref{fig:compare_A2_rhomax} show the numerical results obtained by \texttt{Gmunu}, where the ideal MHD simulation is shown in navy while the resistive simulation with low conductivity $\eta=10^{-6}$ is shown in red.
Since the low resistivity $\eta =10^{-6}$ corresponds to Ohmic decay timescale in the order of $10^5~\rm{ms}$, both of these results is expected to nearly identical in such short timescale.
However, some minor difference is unavoidable because the implementation of the ideal versus resistive MHD are different (see \cite{2021MNRAS.508.2279C, 2022ApJS..261...22C}).

The dashed line in Fig.~\ref{fig:compare_A2_rhomax} represents the ideal magnetohydrodynamical evolution as modelled by \texttt{IllinoisGRMHD}, as detailed in~\cite{2022PhRvL.128f1101T}. 
The oscillation generated by this model has a larger amplitude compared to that produced by \texttt{Gmunu}. 
Several factors may account for this difference: (i) \texttt{Gmunu} utilizes a conformally-flat approximation, whereas \texttt{IllinoisGRMHD} employs a fully general relativistic approach; (ii) \texttt{Gmunu} solves the metric quantities elliptically after introducing the force-free-like atmosphere around the star at the start of the simulation, leading to an initial condition closer to equilibrium, unlike \texttt{IllinoisGRMHD}. 
Despite the variance in oscillation amplitude, the results produced by~\texttt{Gmunu} are generally in agreement with the reference solution from \texttt{IllinoisGRMHD}.


\bibliography{references}{}

\end{document}